\documentclass[a4paper,11pt]{article}
\pdfoutput=1
\usepackage{jcappub}
\usepackage[T1]{fontenc}

\usepackage{graphicx}

\newcommand{\lsim}{\mathrel{\hbox{\rlap{\lower.55ex\hbox{$\sim$}} \kern-.3em \raise.4ex \hbox{$<$}}}}
\newcommand{\gsim}{\mathrel{\hbox{\rlap{\lower.55ex\hbox{$\sim$}} \kern-.3em \raise.4ex \hbox{$>$}}}}

\newcommand{\vp}{\varphi}

\newcommand{\mpl}{m_{\mbox{\tiny{Pl}}}}
\newcommand{\Beq}{\begin{equation}\begin{aligned}}
\newcommand{\Eeq}{\end{aligned}\end{equation}}

\title{GFiRe \fontsize{18}{0}\sc{Gauge Field integrator for Reheating}}

\author[a]{Kaloian D. Lozanov}
\author[b]{and Mustafa A. Amin}

\affiliation[a]{\,Max Planck Institute for Astrophysics, Karl-Schwarzschild-Str.1, 85741 Garching, Germany,}
\affiliation[b]{\,Physics \& Astronomy Department, Rice University, 6100 Main Street, Houston, U.S.A.}

\emailAdd{klozanov@mpa-garching.mpg.de, mustafa.a.amin@rice.edu}

\abstract{We present a new numerical algorithm and code, ${\sf GFiRe}$, for solving the non-linear evolution of Abelian gauge fields coupled to complex scalar fields in homogeneous and isotropic spacetimes. We adopt a hybrid approach to solving the system: the spatial derivatives are discretized using standard Lattice Gauge Field Theory techniques, whereas the time evolution of the fields and scale factor is implemented with explicit, composite, symplectic integrators. An important property of our compound algorithm is that the discretized Gauss constraint is respected exactly, regardless of the order of the symplectic integrator. This remains true even when the background expansion is computed ``self-consistently''; that is, when the expansion history is computed using spatial averaged components of the energy momentum tensor in the simulation volume. Hence, our code can also be used in cases where the fields dominate the energy density of the universe, for example, during reheating after inflation.  

We test the algorithm in scenarios of reheating where the inflaton is a complex scalar field with a potential $\propto(2|\varphi|^2-v^2)^2$ and is coupled to an Abelian gauge field. Tracing the evolution of the system through complex dynamics (including resonant excitation of fields, backreaction, formation of solitons, and changes in the equation of state) in a self-consistently expanding universe, we find the energy conservation violation ($<10^{-4}$) to be very stable and the Gauss constraint violation ($<10^{-6}$) to be dominated by differencing noise.}

\begin{document}
\maketitle
\flushbottom

\section{Introduction}


Reheating after inflation is an integral part of the inflationary scenario (see, for example, \cite{Dolgov:1989us,Traschen:1990sw,Kofman:1994rk,Shtanov:1994ce,Kofman:1997yn,Bassett:2005xm,Allahverdi:2010xz,Frolov:2010sz,Amin:2014eta,Lozanov:2019jxc}). 
Non-nonlinear dynamics during reheating can lead to novel phenomenon such as: (i) changes in the post-inflationary equation of state (e.g., \cite{Podolsky:2005bw,Lozanov:2016hid,Lozanov:2017hjm}), (ii) formation of defects (for example, cosmic strings \cite{Hindmarsh:1994re,Copeland:2009ga}) and solitons (e.g., oscillons, \cite{Copeland:2009ga,Amin:2011hj,Gleiser:2011xj}, and related structures \cite{PhysRevD.66.043505,Amin:2019ums,Niemeyer:2019gab,Musoke:2019ima}), (iii) production of non-gaussian gravitational perturbations (e.g.,~\cite{Chambers:2007se,Bond:2009xx,Imrith:2019njf}), (iv) stochastic gravitational waves (e.g.,~\cite{Khlebnikov:1997di,Easther:2006gt,Dufaux:2010cf,Lozanov:2019ylm,Adshead:2019lbr,Adshead:2019igv,Zhou:2013tsa,Antusch:2016con,Antusch:2017flz,Antusch:2017vga,Liu:2018rrt,Easther:2006vd,GarciaBellido:2007af,Dufaux:2007pt,Dufaux:2008dn,Bartolo:2016ami,Figueroa:2017vfa,Caprini:2018mtu,Bartolo:2018qqn,Armendariz-Picon:2019csc}), and (v) formation of primordial black holes (e.g.,~\cite{1985MNRAS.215..575K,Cotner:2019ykd,Martin:2019nuw,GarciaBellido:1996qt,Green:2000he,Hidalgo:2011fj,Torres-Lomas:2014bua,Suyama:2004mz,Suyama:2006sr,Cotner:2018vug}).  Many non-thermal and nonlinear phenomena during reheating can potentially change abundances of relics such as dark matter and baryon asymmetry (e.g.,~\cite{Garcia:2018wtq,Lozanov:2014zfa,Hertzberg:2013jba,Enqvist:1997si}). 


The details of the nonlinear phenomena are often hard to infer directly from from analytic considerations, making numerical simulations necessary in many cases.  The non-trivial scale dependence arising from non-perturbative dynamics can provide insights into the best observational strategies to probe new physics from this era. To have confidence in new phenomena discovered in the simulations, and their observational implications, the numerical algorithms and simulations must satisfy known analytic constraints and the numerical errors must be under control. It is therefore important to develop techniques for studying the nonlinear evolution of fields as accurately as possible. Moreover, many techniques developed for studying nonlinear dynamics of fields in reheating, can be applied to other eras in the early universe including phase transitions (e.g.~\cite{Hindmarsh:2001vp}), moduli dynamics (e.g~\cite{Giblin:2017wlo,Amin:2019qrx}) and also to the dynamics of axion/axion-like fields (e.g.~\cite{Kolb:1993hw,Kitajima:2018zco,Amin:2019ums,Buschmann:2019icd,Fukunaga:2019unq,Patel:2019isj}) and even in the context of resolving the Hubble tension \cite{Smith:2019ihp}. 

Gauge fields are present in the Standard Model and its extensions, for example the photon, $W$ and $Z$ bosons. They can have highly nonlinear dynamics during reheating, partly due to their Bosonic nature. Their simulations, however, are challenging -- first because of the increase in the number of dynamical degrees of freedom, but more importantly, because of the requirement of preserving certain additional constraints (for example, the Gauss constraint) during the time evolution. In this work, we provide an algorithm and code, ${\sf GFiRe}$ -- {\it a Gauge Field Integrator for Reheating} which is capable of simulating a system of charged scalar fields and Abelian Gauge fields on a lattice in an homogeneous and isotropic, expanding universe. ${\sf GFiRe}$ preserves the Gauss constraint and energy constraint exceptionally well; this remains true even when the expansion is determined {\it self-consistently} by the spatially averaged energy momentum tensor at each time step.  Our time evolution is symplectic and explicit in time. 

Below, we briefly review some relevant literature on nonlinear dynamics of cosmological fields (mostly in the context of reheating), and put our current work in context of the existing literature -- highlighting the new and useful aspects of our present work.

Exploring the nonlinear dynamics of scalar fields during reheating is not new. Soon after the importance of non-adiabatic particle production in the oscillating inflaton background was appreciated (a phenomenon known as preheating \cite{Traschen:1990sw,Kofman:1994rk,Shtanov:1994ce,Kofman:1997yn}), it was realized that the subsequent backreaction can lead to a complex, non-linear evolution of the combined inflaton-daughter fields system. For some of the earliest scalar field simulations in the reheating context, see \cite{Khlebnikov:1996mc,Prokopec:1996rr}. Since then, a number of dedicated lattice codes have been developed for scalar field models. Most of them are based on finite-differencing techniques, e.g., {\sc LatticeEasy} \cite{Felder:2000hq}, {\sc Defrost} \cite{Frolov:2008hy}, {\sc HLattice} \cite{Huang:2011gf}, {\sc PyCool} \cite{Sainio:2012mw}, {\sc Gabe} \cite{Child:2013ria}, but there are also pseudo-spectral ones, e.g., {\sc PSpectRe} \cite{Easther:2010qz} and {\sc Stella} \cite{Amin:2018xfe}.  Usually, the codes solve the evolution of a system of spatially inhomogeneous, (self-)interacting scalar-field(s) in a spatially homogeneous and isotropic spacetime. In many cases, the expansion rate is determined self-consistently from the acceleration equation, with appropriate preservation of the energy constraint (i.e., the Friedmann equation is satisfied).\footnote{In most of the mentioned codes, metric perturbations are typically sourced passively, i.e., they do not backreact on the dynamics of the scalar fields. Some exceptions are HLATTICE and the recent GABERel \cite{Giblin:2019nuv}, which allow for the full general relativistic evolution in reheating scalar field models. Non-minimal couplings to gravity (but with passively evolved perturbations) have also been considered recently \cite{Rubio:2019ypq,Nguyen:2019kbm,Crespo:2019src,Crespo:2019mmh}. In the non-relativistic limit of the fields, there exist works that carry out lattice simulations with backreaction of the scalar gravitational perturbations included in both expanding and non-expanding spacetimes (see for example \cite{Schive:2014dra,Schwabe:2016rze,Edwards:2018ccc,Amin:2019ums}.)} 

Beyond scalar fields, lattice simulations of gauge fields have also been carried out. These scenarios can be split into two groups depending on the nature of the scalar coupled to the gauge fields: (i) the scalar is uncharged ($\chi$), and (ii) the scalar is charged ($\varphi$). 

For an uncharged scalar + Abelian gauge fields systems, Lagrangians which include couplings of the form $\chi F\tilde{F}$ \cite{Adshead:2015pva,Adshead:2016iae,Adshead:2018doq,Cuissa:2018oiw,Figueroa:2017qmv,Figueroa:2019jsi,Braden:2010wd}, and $f(\chi)F^2$ \cite{Deskins:2013lfx} have been simulated in the context of reheating. For non-Abelian fields,  $\mathcal{L}\supset f(\chi)\rm{tr\boldsymbol F^2}$ \cite{Adshead:2017xll} has been explored. In \cite{Cuissa:2018oiw,Figueroa:2017qmv,Figueroa:2019jsi}, gauge invariance at the level of the spatially discretized action is guaranteed by the use of slightly modified {\it Link Variables} \cite{smit2002introduction}, and the Gauss constraint is preserved during the time evolution. The expansion of the universe is obtained from the spatially averaged energy density. However, their algorithm (which works in the temporal gauge), required that the integration in time is {\it implicit}. On the other hand, the authors in \cite{Adshead:2015pva,Adshead:2016iae,Adshead:2018doq} use an explicit time evolution scheme. A Lorenz gauge is used, where all gauge degrees of freedom are made dynamical. Gauge invariance at the level of the spatially discretized action, and the preservation of the Lorenz gauge condition by time evolution is not guaranteed, however, the Lorenz gauge condition is carefully checked a-posteriori after solving the equations of motion on the lattice. The expansion of the universe is once again determined self-consistently.

For the case of {\it charged} scalar and gauge fields, lattice simulations with Abelian \cite{Rajantie:2000fd,Dufaux:2010cf,Figueroa:2015rqa,Figueroa:2016ojl,Figueroa:2016dsc} and non-Abelian \cite{GarciaBellido:2003wd,Graham:2006vy,DiazGil:2007dy,Graham:2007ds,DiazGil:2008tf,Enqvist:2015sua,Mou:2017zwe,Tranberg:2017lrx} gauge fields have been carried  where the spatially discretized action, and the time evolution respects the Gauss constraint by using standard Link Variables. However, these simulations assume that the scalar and gauge fields are {\it spectators} (i.e., energetically subdominant). The energy constraint in an expanding universe is not relevant in these studies since they assume that the Friedmann-Robertson-Walker (FRW) expansion is determined by an unrelated component of the Universe, e.g., a dark matter component or a thermal bath. The evolution of the scale factor is set to a fixed power-law, $a(\tau)\propto \tau^p$ (for example, with $p=1$ for radiation domination). 

Our present algorithm and code, {\sf GFiRe}, includes and adds to the desirable features that appear separately in the above mentioned works. In particular, some of its attractive features include: 
\begin{itemize}
\item Our discretized action respects a residual gauge invariance related to time-independent gauge transformations (we work in a gauge where the time component of the gauge field is set to zero). This is achieved by using Link Variables for the spatial discretization. The Gauss constraint follows naturally from residual symmetries of the spatially discretized action.
\item  Our time evolution scheme is explicit in time, and also symplectic. This scheme treats the time evolution of the scale factor (whose equations of motion involve spatial averaging of the energy momentum tensor) in the same way it treats the time evolution of the fields.\footnote{In the context of reheating, symplectic integrator techniques \cite{Yoshida:1990zz} for the time integration were first employed by the scalar field codes HLATTICE, DEFROST and PyCOOL. For some scalar field models with non-canonical kinetic terms, symplectic integrators have been employed by \cite{Krajewski:2018moi}. } 
\item We show that our symplectic time evolution scheme (of the scale factor and fields) preserves the Gauss constraint.  We show explicitly that the Gauss constraint is respected exactly by non-linear symplectic integrators of arbitrary order. This result allows us to take advantage of the properties of high order symplectic integrators, for example, excellent stability and conservation for relatively large time steps without the need to worry about spurious gauge modes.
\end{itemize}
We note that when the charged scalar is the inflaton, or more generally, when the fields dominate the energy density of the universe,  solving accurately for the FRW expansion and the equation-of-state can be important for the observational consequences of reheating \cite{Lozanov:2017hjm,Lozanov:2016hid, Adshead:2019lbr,Adshead:2019igv}. Such calculations can substantially reduce uncertainties in inflationary observables such as the tensor-to-scalar ratio, $r$, and the spectral index $n_{\rm s}$ \cite{Lozanov:2017hjm}. Self-consistent expansion with good energy-constraint preservation can also be important for studies of non-gaussianity \cite{Chambers:2007se,Bond:2009xx}.

The paper is organized as follows. In Section \ref{sec:ScalarQED} we set the scene for the symplectic integrators. After reviewing the generic equations of motion of the model, we rewrite the system in a Hamiltonian form and discretize it on a spatial lattice. In Section \ref{sec:GFiReImplementation} we introduce the symplectic integrator algorithm and apply it to the Hamiltonian lattice system. We present the results from two test studies with our new algorithm in Section \ref{sec:NumStudies}. Section \ref{sec:Disc} is devoted to concluding remarks and a discussion. Throughout the paper we use natural units in which $c=\hbar=1$ and the reduced Planck mass is $\mpl=1/\sqrt{8\pi G}$. For the metric, we use the $+---$ signature.

\section{Scalar Electrodynamics}
\label{sec:ScalarQED}

We consider classical scalar Electrodynamics, also known as the Abelian-Higgs or the Ginzburg-Landau model, with a general Higgs potential. The action for this theory is
\Beq
\label{eq:SAHFRW}
S=\int d^4x\sqrt{-g}\mathcal{L}=\int d^4x\sqrt{-g}\left[-\frac{\mpl^2}{2}R+|\mathcal{D}\varphi|^2-\frac{1}{4e^2}F_{\mu\nu}F^{\mu\nu}-\mathcal{V}(|\varphi|)\right]\,,
\Eeq
where $\vp$ is a complex valued scalar field, $g$ is the determinant of the metric, and $R$ is the Ricci scalar. The covariant derivative and the field tensor are
\Beq
\mathcal{D}_\mu\equiv\partial_{\mu}+iA_{\mu}\,,\quad F_{\mu\nu}\equiv\mathcal{D}_{\mu}A_{\nu}-\mathcal{D}_{\nu}A_{\mu}=\partial_{\mu}A_{\nu}-\partial_{\nu}A_{\mu}\,,
\Eeq
with $A_\mu$ being a Lorentz vector field. The Lagrangian density in Eq. \eqref{eq:SAHFRW} is invariant under the local (gauge) symmetry transformation
\Beq
\varphi\rightarrow\exp[-i\alpha(x^\nu)]\varphi\,,\quad A_\mu\rightarrow A_\mu+\partial_\mu\alpha(x^\nu)\,,
\Eeq
where $\alpha(x^\nu)$ is an arbitrary, real valued function of spacetime. 

Varying the action with respect to $g_{\mu\nu}$ yields the Einstein Equations:
\Beq
\label{eq:EinsEq}
&\mathcal{G}_{\mu\nu}=R_{\mu\nu}-\frac{1}{2}g_{\mu\nu}R=\frac{T_{\mu\nu}}{\mpl^2}\,,\\
&T_{\mu\nu}=2\left(\mathcal{D}_{(\mu}\varphi\right)^*\mathcal{D}_{\nu)}\varphi-e^{-2}F_{\mu\alpha}F_{\nu}^{\,\,\,\alpha}-g_{\mu\nu}\left[\left(\mathcal{D}^{\alpha}\varphi\right)^*\mathcal{D}_{\alpha}\varphi-\mathcal{V}-\frac{F_{\alpha\beta}F^{\alpha\beta}}{4e^2}\right]\,,
\Eeq
where $\mathcal{G}_{\mu\nu}$ is the Einstein tensor, $R_{\mu\nu}$ is the Ricci tensor and $T_{\mu\nu}$ is the energy momentum tensor that includes contributions from $\vp$ and $A_\mu$. Varying the action \eqref{eq:SAHFRW} with respect to $\vp$ and $A_\mu$ yields the Euler Lagrange equations of motion for $\vp$ and $A_\mu$:
\Beq
\label{eq:EoMLagr}
&\mathcal{D}_{\mu}\mathcal{D}^{\mu}\varphi+\frac{\partial\mathcal{V}}{\partial\varphi^*}=0\,,\\
&\nabla_{\mu}F^{\mu\nu}-2e^2\Im\left[\varphi\left(\mathcal{D}^{\nu}\varphi\right)^*\right]=0\,.\\
\Eeq
In particular, the $0$ component of the second equation was obtained by varying the action with respect to $A_0$. This is the Gauss constraint.  For future convenience we define
\Beq
\mathcal{C}_{\rm G}\equiv\sqrt{-g}\left\{\nabla_{\mu}F^{\mu0}-2e^2\Im\left[\varphi\left(\mathcal{D}^{0}\varphi\right)^*\right]\right\}=0\,.
\Eeq
Note that $\mathcal{C}_{\rm G}=0$ follows from  the $0$th component of the second line of \eqref{eq:EoMLagr}. Moreover, even without using $\mathcal{C}_{\rm G}=0$, the equation of motion for $\vp$ and spatial part of the second line of \eqref{eq:EoMLagr}, implies 
\Beq
\partial_0 \mathcal{C}_{\rm G}=0\,.
\Eeq

What if we had fixed $A_0=0$ at the level of the action? In this case, the Euler-Lagrange equations do not yield the Gauss constraint ($\mathcal{C}_{\rm G}=0$). Nevertheless, there is an alternate route to arrive at the correct expression for $\mathcal{C}_{\rm G}$, and also show that $\partial_0 \mathcal{C}_{\rm G}=0$. In this route, we can make use of the residual symmetry $A_i \rightarrow A_i+\partial_i\alpha({\bf x})$ and $\vp\rightarrow \vp e^{-i\alpha({\bf x})}$ of the action (left over even after $A_0=0$ is fixed) to arrive at an infinite number of Noether charges $q({\bf x})=\mathcal{C}_{\rm G}$ with $\partial_0\mathcal{C}_{\rm G}=0$ on the equations of motion (see Ch. 10 in \cite{Weinberg:2012pjx}). The key thing we miss out on is that we now need to set $\mathcal{C}_{\rm G}=0$ ``by hand'' at some initial time.

We now restrict ourselves to a homogeneous and isotropic universe where
\Beq
\label{eq:FRW}
g_{\mu\nu}=a^2(\tau)\eta_{\mu\nu}\,,\quad{\rm with}\qquad\eta_{\mu\nu}={\rm diag}[1,-1,-1,-1]\,.
\Eeq
Here, $\tau$ is conformal time, and $a(\tau)$ is the scale factor. Furthermore, we chose to work in the temporal gauge with $A_0=0$. In this case, the equations of motion \eqref{eq:EoMLagr} become
\Beq
\label{eq:FRWEulLagr}
&\partial_{\tau}^2\varphi+2\mathcal{H}\partial_{\tau}\varphi-\Delta\varphi-2iA_{j}\partial_{j}\varphi-i\varphi\partial_{j}A_{j}+A_{j}A_{j}\varphi+a^2\frac{\partial\mathcal{V}}{\partial\varphi^*}=0\,,\\
&\partial_{\tau}^2A_{j}-\Delta A_{j}+\partial_{j}\partial_{i}A_i-2e^2a^2\Im\left[\varphi\left(\mathcal{D}_{j}\varphi\right)^*\right]=0\,,\\ \\
&\mathcal{C}_{\rm G}= \partial_{\tau}\partial_{j}A_j-2e^2a^2\Im\left[\varphi\left(\mathcal{D}_{0}\varphi\right)^*\right]=0\,,\\
\Eeq
where the last equation is the Gauss constraint. The trace, and the $00$ component of the Einstein equations \eqref{eq:EinsEq} (assuming FRW spacetimes) yield\footnote{For reference,
\Beq
R_{00}=3\left[\mathcal{H}^2-\frac{a''}{a}\right]\,,\qquad R_{ij}=\left[\mathcal{H}^2+\frac{a''}{a}\right]\delta_{ij}\,,\qquad R=-6\frac{a''}{a^3}\,.
\Eeq}:
\Beq
\label{eq:FRWRych}
\frac{a''}{a}&=\frac{1}{2m_{\rm Pl}^2}a^2\left(\rho-3p\right)\,,\\
\mathcal{C}_{\mathcal{E}}&\equiv\mathcal{H}^2-\frac{1}{3m_{\rm Pl}^2}a^2\rho=0\,,
\Eeq
where $\mathcal{H}\equiv a'/a$, and we have defined the spatially averaged density and pressure as $\rho\equiv\langle T_{00}\rangle/a^2$ and $p\equiv \delta^{ij}\langle T_{ij}\rangle/(3a^2)$. In analogy with the Gauss constraint, the energy constraint $\mathcal{C}_{\mathcal{E}}=0$. 

There is an alternative treatment for arriving at the evolution equation for the scale factor by first approximating the spacetime to be FRW at the level of the action (rather than at the level of the equations of motion). Explicitly,  we first integrate the FRW Einstein-Hilbert term in the action by parts
and then extremize the action with respect to the scale factor, to arrive at the first equation in \eqref{eq:FRWRych}. Note that this time, the spatial averaging follows directly from extremizing the action.

In summary, we have shown that even if $A_0=0$ is enforced at the level of the action, we arrive at $\partial_0\mathcal{C}_{\rm G}=0$ via the equations of motion. Separately, assuming an FRW universe at the level of the action yields the same equation of motion for the scale factor as the one obtained by imposing the FRW symmetries at the level of the equations of motion. Below, we will arrive at the same results in the Hamiltonian formulation.

\subsection{Hamiltonian formalism in $A_0=0$ gauge and FRW spacetimes}

Our algorithm and code is developed from a Hamiltonian formulation of the equations of motion, which we derive in this section. Here, we will work directly in the $A_0=0$ gauge, and assume an FRW universe from the beginning. We will also be laboriously explicit in our expressions since it will ease the route to their corresponding discretized versions.

\subsubsection{Action and Hamiltonian}

The action \eqref{eq:SAHFRW} becomes
\Beq
\label{eq:SA0}
S_{0}&=\int d\tau \,\,L_{0}(\tau)=\int d^4x\,\,\mathcal{L}_{0}(\tau,{\bf x})\\
  &=\int d^4x \left[-3\mpl^2a'^2+\frac{(a\varphi_1')^2}{2}+\frac{(a\varphi_2')^2}{2}-a^2|\boldsymbol{\mathcal{D}}\varphi|^2-a^4V+\frac{\textbf{A}\!'^2}{2e^2}-\frac{(\boldsymbol{\nabla}\times\textbf{A})^2}{2e^2}\right]\,,
\Eeq
where the $0$ subscript reminds us that we are in the temporal gauge and in an FRW universe. The complex field $\vp$ is written in terms of two real fields, 
\Beq
\varphi=\frac{1}{\sqrt{2}}(\varphi_1+i\varphi_2)\,,\quad \textrm{with}\quad\mathcal{V}(|\varphi|)=V\left(\sqrt{\varphi_1^2+\varphi_2^2}\right)\,.
\Eeq
Our gauge choice allows us to apply the Hamiltonian formalism readily, since all variables
\Beq
\varphi_1(\tau,\textbf{x})\,,\quad\varphi_2(\tau,\textbf{x})\,,\quad\textbf{A}(\tau,\textbf{x})=[A_1(\tau,\textbf{x})\,,A_2(\tau,\textbf{x})\,,A_3(\tau,\textbf{x})]^T\,\quad \textrm{and}\quad a(\tau)\,
\Eeq
are dynamical. 
Their conjugate momenta are
\Beq
\pi_1(\tau,\textbf{x})=\frac{\partial}{{\partial\varphi_1}'}\mathcal{L}_{0}\,,\quad\pi_2(\tau,\textbf{x})=\frac{\partial}{{\partial\varphi_2}'}\mathcal{L}_{0}\,,\quad\boldsymbol{\pi}_A(\tau,\textbf{x})=\frac{\partial}{\partial\textbf{A}\!'}\mathcal{L}_{0}\,,\quad p_a(\tau)=\frac{\partial}{{\partial a}'}L_{0}\,,
\Eeq
which reduce to\footnote{The Einstein-Hilbert term in the action, see Eqs.~(\ref{eq:SA0}), in the FRW approximation involves a spatially-independent $3\mpl^2a'(\tau)^2$ factor integrated over space. The factor can be moved in front of the 3-dimensional integral, yielding a formally divergent 3-volume integral over unity. The latter can be written as $\int d^3x=(2\pi)^3\delta^3(0)$, after applying the Fourier representation of the Dirac delta function.}
\Beq
\label{eq:ConjMomenta}
\pi_1=a^2\varphi_1'\,,\quad \pi_2=a^2\varphi_2'\,,\quad \boldsymbol{\pi}_{\!A}=e^{-2}\textbf{A}\!'\,,\quad p_a=-6\mpl^2a'(2\pi)^3\delta^{3}(0)\,.
\Eeq
The Hamiltonian is found from the Legendre transformation
\Beq
\label{eq:HamiltonianA0FRW}
H_{0}(\tau)&=p_a a'+\int d^3x\left[\pi_1\varphi_1'+\pi_2\varphi_2'+\boldsymbol{\pi}_A\cdot\textbf{A}'\right]-\int d^3x\mathcal{L}_{0}\\
&=-\frac{1}{(2\pi)^3\delta^{3}(0)}\frac{p_a^2}{12\mpl^2}+\int d^3x\Big[\frac{\pi_1^2}{2a^2}+\frac{\pi_2^2}{2a^2}+\frac{(e\boldsymbol{\pi}_A)^2}{2}\\
&\qquad\qquad\qquad\qquad\qquad\qquad\qquad\qquad+a^2|\boldsymbol{\mathcal{D}}\varphi|^2+a^4V+\frac{(\boldsymbol{\nabla}\times\textbf{A})^2}{2e^2}\Big]\,.
\Eeq

\subsubsection{Hamilton's equations}

The action in Eq. \eqref{eq:SA0} is extremised when the Hamilton's equations of motion are satisfied. To write the equations of motion compactly, we use Poisson brackets $\{\cdot,\cdot\}_{\rm P}$, defined in our case as 
\Beq
\label{eq:PoissonBracket}
\{\mathcal{A(\tau,\textbf{x}')},\mathcal{B(\tau,\textbf{x}'')}\}_{\rm P}\equiv&\frac{\partial\mathcal{A}(\tau,\textbf{x}')}{\partial a(\tau)}\frac{\partial \mathcal{B(\tau,\textbf{x}'')}}{\partial p_a(\tau)}-\frac{\partial \mathcal{B}(\tau,\textbf{x}'')}{\partial a(\tau)}\frac{\partial \mathcal{A}(\tau,\textbf{x}')}{\partial p_a(\tau)}\\
&+\int d^3x\sum_i\left[\frac{\delta\mathcal{A}(\tau,\textbf{x}')}{\delta \Psi_i(\tau,\textbf{x})}\frac{\delta \mathcal{B}(\tau,\textbf{x}'')}{\delta \Pi_i(\tau,\textbf{x})}-\frac{\delta \mathcal{B}(\tau,\textbf{x}'')}{\delta \Psi_i(\tau,\textbf{x})}\frac{\delta\mathcal{A}(\tau,\textbf{x}')}{\delta \Pi_i(\tau,\textbf{x})}\right]\,,
\Eeq
where $i$ runs from $1$ to $5$, $\Psi=[\varphi_1,\varphi_2,A_1,A_2,A_3]^T$, $\Pi=[\pi_1,\pi_2,\pi_{A_1},\pi_{A_2},\pi_{A_3}]^T$. The $\mathcal{A}$ and $\mathcal{B}$ are functions (not functionals) of the fields, fields momenta and their spatial derivaties, i.e.,
\Beq
\mathcal{A}(\tau,\textbf{x})&\equiv\mathcal{A}(a(\tau),\Psi_i(\tau,\textbf{x}),\nabla\Psi_i(\tau,\textbf{x}),p_a(\tau),\Pi_i(\tau,\textbf{x}),\nabla\Pi_i(\tau,\textbf{x}))\,,\\
\mathcal{B}(\tau,\textbf{x})&\equiv\mathcal{B}(a(\tau),\Psi_i(\tau,\textbf{x}),\nabla\Psi_i(\tau,\textbf{x}),p_a(\tau),\Pi_i(\tau,\textbf{x}),\nabla\Pi_i(\tau,\textbf{x}))\,,
\Eeq
and we also define
\Beq
\frac{\delta\mathcal{A}(\tau,\textbf{x}')}{\delta \Psi_i(\tau,\textbf{x})}\equiv\left[\frac{\partial\mathcal{A}(\tau,\textbf{x})}{\partial \Psi_i(\tau,\textbf{x})}-\partial_j\frac{\partial\mathcal{A}(\tau,\textbf{x})}{\partial \partial_j\Psi_i(\tau,\textbf{x})}\right]\delta({\bf x}'-{\bf x})\,,
\Eeq
etc. We can now express Hamilton's equations as
\Beq
\label{eq:HamEq}
\partial_{\tau}a&=\left\{a,H_{0}\right\}_{\rm P}\,,\\
\partial_{\tau}p_a&=\left\{p_a,H_{0}\right\}_{\rm P}\,,\\
\partial_{\tau}\Psi_i&=\left\{\Psi_i,H_{0}\right\}_{\rm P}\,,\\
\partial_{\tau}\Pi_i&=\left\{\Pi_i,H_{0}\right\}_{\rm P}\,,
\Eeq
and note that the Hamiltonian from Eq. \eqref{eq:HamiltonianA0FRW} can be rewritten as
\Beq
H_{0}&=\int d^3x \mathcal{H}_{0}(\textbf{x})\\
&=\int d^3x\Big[-\frac{1}{((2\pi)^3\delta^{3}(0))^2}\frac{p_a^2}{12\mpl^2}+\frac{\pi_1^2}{2a^2}+\frac{\pi_2^2}{2a^2}+\frac{(e\boldsymbol{\pi}_A)^2}{2}\\
&\qquad\qquad\qquad\qquad\qquad\qquad\qquad\qquad+a^2|\boldsymbol{\mathcal{D}}\varphi|^2+a^4V+\frac{(\boldsymbol{\nabla}\times\textbf{A})^2}{2e^2}\Big]\,.
\Eeq
The first and third lines in Eq. \eqref{eq:HamEq} are the definitions of the conjugate momenta given in Eq. \eqref{eq:ConjMomenta}. The second line in Eq. \eqref{eq:HamEq} yields the evolution of the scale factor and is identical to the first equation in \eqref{eq:FRWRych}, i.e., the trace of the Einstein equations (note that the spatial averaging follows directly from the Hamilton's equations). The last line in Eq. \eqref{eq:HamEq} gives the dynamical equations of motion for the real and imaginary parts of the Higgs and the spatial components of the gauge field, i.e., the first two lines in Eq. \eqref{eq:FRWEulLagr} (with $A_0=0$). 

\subsubsection{Gauss constraint}
Where is the Gauss constraint in this Hamiltonian prescription? As discussed earlier, the residual symmetry $A_i \rightarrow A_i+\partial_i\alpha({\bf x})$ and $\vp\rightarrow \vp e^{-i\alpha({\bf x})}$ of the action \eqref{eq:SA0} left over after setting $A_0=0$, leads to an infinite number of Noether charges which essentially leads us to the Gauss constraint \cite{Weinberg:2012pjx}. Explicitly, the infinitesimal symmetry transformations are 
\Beq
\delta\varphi_1=\varphi_2\alpha\,,\quad\delta\varphi_2=-\varphi_1\alpha\,,\quad \delta\textbf{A}=\boldsymbol{\nabla}\alpha\,,
\Eeq
where $|\alpha(\textbf{x})|\ll1$.  The conserved Noether current is given by
\Beq
j^\mu=\sum_i\frac{\partial\mathcal{L}_{A_0}^{{}^{\rm FRW}}}{\partial_{\mu}\Psi_i}\delta \Psi_i\,.
\Eeq
The corresponding constant charge is
\Beq
Q=\int d^3x\,j^0&=\int d^3x \left[\boldsymbol{\pi}_{\!A}\cdot\boldsymbol{\nabla}\alpha+(\pi_1\varphi_2-\pi_2\varphi_1)\alpha\right]\\
                             &=\oint\alpha\boldsymbol{\pi}_{\!A}\cdot\textbf{dS}-\int d^3x\,\alpha\left[\boldsymbol{\nabla}\cdot\boldsymbol{\pi}_{\!A}-(\pi_1\varphi_2-\pi_2\varphi_1)\right]\,.
\Eeq
Since $Q=\text{const}.$ for arbitrary $\alpha(\textbf{x})$, then
\Beq
\label{eq:CGauss}
e^{-2}\mathcal{C}_{\rm G}(\tau,\textbf{x})=\boldsymbol{\nabla}\cdot\boldsymbol{\pi}_{\!A}(\tau,\textbf{x})-\left[\pi_1(\tau,\textbf{x})\varphi_2(\tau,\textbf{x})-\pi_2(\tau,\textbf{x})\varphi_1(\tau,\textbf{x})\right]=\text{const}\,, 
\Eeq
provided $\lim_{|\boldsymbol{x}|\rightarrow\infty} \alpha(\textbf{x})\boldsymbol{\pi}_{\!A}(\tau,\textbf{x})= \boldsymbol{0}$, which must hold, since $\alpha$ and $\boldsymbol{\pi}_A$ vanish separately at infinity -- the former according to the definition of a gauge transformation and the latter trivially. Hence, the gauge fixed action, \eqref{eq:SA0}, enforces $\partial_{\tau}\mathcal{C}_{\rm G}=0$ and so do its Hamiltonian equations. 

As a final remark we recall that when the Gauss constraint was derived in the case where the gauge was fixed after deriving the equations of motion, we ended up with $\mathcal{C}_{\rm G}(\tau,{\bf x})=0$. In our Hamiltonian formulation, we arrived at $\partial_\tau \mathcal{C}_{\rm G}(\tau,{\bf x})=0$. We can readily restrict ourselves to $\mathcal{C}_{\rm G}(\tau,{\bf x})=0$ at all times in our Hamiltonian formulation by simply imposing the Gauss constraint on the initial time slice, $\mathcal{C}_{\rm G}(\tau_{\rm in},\textbf{x})=0$, and then use Hamiltonian equations to get $\mathcal{C}_{\rm G}(\tau,\textbf{x})=0$.

\subsection{Hamiltonian formalism in $A_0=0$ gauge on a comoving lattice}

\begin{figure}
 \centering
 \includegraphics[width=6.0in]{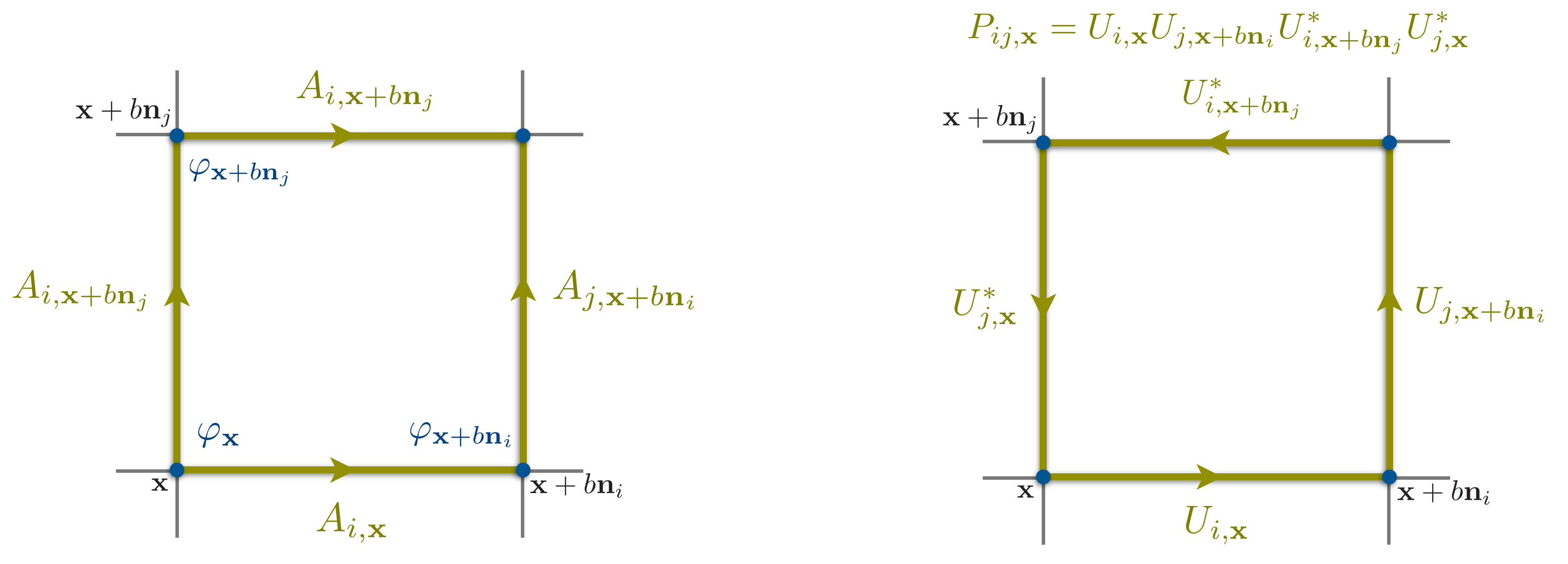}
 \caption{A visual representation of our spatial discretization scheme. Here $b=\Delta x$. The fields $\vp$ live on the lattice nodes, whereas the gauge fields ${\bf A}$ live on the links.}
 \label{fig:Plaquette}
\end{figure}

The next step towards solving the Abelian-Higgs model numerically is the discretization of the fields on three dimensional spatial lattice. We assume a comoving cubic grid with periodic boundary conditions and $N^3$ points. The comoving length of the edge of the unit cell is equal to 
\Beq
b\equiv\Delta x.
\Eeq 
The scalar fields, $\{\varphi_{1,\textbf{x}},\varphi_{2,\textbf{x}}\}$, are defined on the lattice points with {\it discrete} spatial coordinates $\textbf{x}$ and the gauge fields, $A_{j,\textbf{x}}$, are defined on the lattice links, connecting the adjacent points $\textbf{x}$ and $\textbf{x}+\textbf{n}_j$. The unit vectors are
\Beq
\textbf{n}_1=[1,0,0]^T\,,\quad\textbf{n}_2=[0,1,0]^T\,,\quad\textbf{n}_3=[0,0,1]^T\,.
\Eeq
The standard {\it links}, $U_{j,\textbf{x}}$, and {\it plaquettes}, $ P_{ij,\textbf{x}}$, are \cite{smit2002introduction} 
\Beq
\label{eq:LinksPlaquettes}
U_{j,\textbf{x}}=\exp[ibA_{j,\textbf{x}}]\,,\quad P_{ij,\textbf{x}}=U_{i,\textbf{x}}U_{j,\textbf{x}+\textbf{n}_ib}U_{i,\textbf{x}+\textbf{n}_jb}^*U_{j,\textbf{x}}^*\,,\quad P_{ij,\textbf{x}}=P_{ji,\textbf{x}}^*\,.
\Eeq
See Fig.~\ref{fig:Plaquette} for a visual representation. We can now use these variables to define a lattice action in the $A_0=0$ gauge, which still has a residual time-independent gauge symmetry. Similarly to the continuous case from the previous section, we will show that in the Hamiltonian formalism, the lattice version of the Gauss constraint is recovered from the conserved charges corresponding to the residual gauge invariance of the lattice action.

\subsubsection{Lattice action and Hamiltonian}

The links and the plaquettes from Eq. \eqref{eq:LinksPlaquettes} can be used to define spatial derivates on the lattice (no summation is assumed over repeated indices)
\Beq
(\mathcal{D}_j\varphi)_{(\textbf{x})}\rightarrow \frac{U_{j,\textbf{x}}\varphi_{\textbf{x}+\textbf{n}_jb}-\varphi_{\textbf{x}}}{b}\,,\quad (F_{ij}F^{ij})_{(\textbf{x})}\rightarrow \frac{2-P_{ij,\textbf{x}}-P_{ji,\textbf{x}}}{b^4}\,.
\Eeq
Hence, the lattice version of the continuous action from Eq. \eqref{eq:SA0} becomes 
\Beq
\label{eq:SA0lat}
S_{0}^{\rm latt}&=\int d\tau\,L_{0}^{\rm latt}\\
&=\int d\tau\Bigg\{-3\mpl^2a'^2N^3+\sum_{\textbf{x}}^{N^3}\Bigg(\frac{(a\varphi_{1,\textbf{x}}')^2}{2}+\frac{(a\varphi_{2,\textbf{x}}')^2}{2}-a^4V_{\textbf{x}}+\frac{1}{2e^2}\sum_j^{1,2,3}A_{j,\textbf{x}}'^2\\&\qquad-\frac{1}{2e^2b^4}\sum_{i,j}^{1,2,3}\left[1-P_{ij,\textbf{x}}\right]-\frac{a^2}{2b^2}\sum_j^{1,2,3}\bigg[(\varphi_{1,\textbf{x}+\textbf{n}_jb}^2+\varphi_{2,\textbf{x}+\textbf{n}_jb}^2)+(\varphi_{1,\textbf{x}}^2+\varphi_{2,\textbf{x}}^2)\\
&\qquad-\left((\varphi_{1,\textbf{x}}-i\varphi_{2,\textbf{x}})U_{j,\textbf{x}}(\varphi_{1,\textbf{x}+\textbf{n}_jb}+i\varphi_{2,\textbf{x}+\textbf{n}_jb})+{\rm c.c.}\right) \bigg]\Bigg)\Bigg\}\,,
\Eeq
where ``c.c'' refers to complex conjugate. Notice that the three-dimensional integral over space has become a summation over the lattice points $\int d^3x\rightarrow\sum_{\textbf{x}}^{N^3} b^3$, 
and that the action has been rescaled by a factor of $b^3$ -- a rescaling of the total action has no effect on the Euler-Lagrange or Hamilton's equations of motions. The continuous system of five fields capturing infinitely many degrees of freedom and the scale factor has now become one having $5\times N^3+1$ generalized coordinates, with conjugate momenta (c.f. \eqref{eq:ConjMomenta})
\Beq
\label{eq:LattMomenta}
\pi_{1,\textbf{x}}=a^2\varphi_{1,\textbf{x}}'\,,\quad \pi_{2,\textbf{x}}=a^2\varphi_{2,\textbf{x}}'\,,\quad\pi_{\!Aj,\textbf{x}}=e^{-2}A_{j,\textbf{x}}'\,,\quad\pi_a=-6\mpl^2a'N^3\,.
\Eeq
The Legendre transformation yields the lattice Hamiltonian (c.f. \eqref{eq:HamiltonianA0FRW})
\Beq
\label{eq:HamiltonianA0FRWlatt}
H_{0}^{\rm latt}(\tau)&=\pi_a a'+\sum_{\textbf{x}}^{N^3}\left[\pi_1\varphi_{1,\textbf{x}}'+\pi_2\varphi_{2,\textbf{x}}'+\sum_j^{1,2,3}\pi_{\!Aj,\textbf{x}}A_{j,\textbf{x}}'\right]-L_{0}^{\rm latt}(\tau)\\
&=-\frac{\pi_a^2}{12\mpl^2N^3}+\sum_{\textbf{x}}^{N^3}\Bigg\{\frac{\pi_{1,\textbf{x}}^2}{2a^2}+\frac{\pi_{2,\textbf{x}}^2}{2a^2}+e^2\sum_j^{1,2,3}\frac{\pi_{Aj,\textbf{x}}^2}{2}+\frac{1}{2e^2b^4}\sum_{i,j}^{1,2,3}\left[1-P_{ij,\textbf{x}}\right]\\
&\qquad\qquad\qquad+a^4V_{\textbf{x}}+\frac{a^2}{2b^2}\sum_j^{1,2,3}\bigg[(\varphi_{1,\textbf{x}+\textbf{n}_jb}^2+\varphi_{2,\textbf{x}+\textbf{n}_jb}^2)+(\varphi_{1,\textbf{x}}^2+\varphi_{2,\textbf{x}}^2)\\
&\qquad\qquad\qquad-\bigg((\varphi_{1,\textbf{x}}-i\varphi_{2,\textbf{x}})U_{j,\textbf{x}}(\varphi_{1,\textbf{x}+\textbf{n}_jb}+i\varphi_{2,\textbf{x}+\textbf{n}_jb})+c.c.\bigg) \bigg]\Bigg\}\,.
\Eeq

\subsubsection{Hamilton's equations}

To extremise the lattice action in Eq. \eqref{eq:SA0lat} we need to use the corresponding Hamilton's equations. We can again write them concisely in terms of the lattice Poisson brackets (c.f. \eqref{eq:PoissonBracket})
\Beq
\{\mathcal{A}_{\textbf{x}'},\mathcal{B}_{\textbf{x}''}\}^{\rm latt}_{\rm P}\equiv\frac{\partial\mathcal{A}_{\textbf{x}'}}{\partial a}\frac{\partial \mathcal{B}_{\textbf{x}''}}{\partial \pi_a}-\frac{\partial \mathcal{B}_{\textbf{x}''}}{\partial a}\frac{\partial\mathcal{A}_{\textbf{x}'}}{\partial \pi_a}+\sum_{\textbf{x}}^{N^3}\sum_{i=1}^5\left[\frac{\partial\mathcal{A}_{\textbf{x}'}}{\partial Q_{\textbf{x},i}}\frac{\partial \mathcal{B}_{\textbf{x}''}}{\partial P_{\textbf{x},i}}-\frac{\partial \mathcal{B}_{\textbf{x}''}}{\partial Q_{\textbf{x},i}}\frac{\partial\mathcal{A}_{\textbf{x}'}}{\partial P_{\textbf{x},i}}\right]\,,
\Eeq
where $Q_{\textbf{x}}=[\varphi_{1,\textbf{x}},\varphi_{2,\textbf{x}},A_{1,\textbf{x}},A_{2,\textbf{x}},A_{3,\textbf{x}}]^T$, $P_{\textbf{x}}=[\pi_{1,\textbf{x}},\pi_{2,\textbf{x}},\pi_{A1,\textbf{x}},\pi_{A2,\textbf{x}},\pi_{A3,\textbf{x}}]^T$ and all quantities are evaluated at equal times. $\mathcal{A}$ and $\mathcal{B}$ are functions of the lattice fields and lattice momenta, i.e.,
\Beq
\mathcal{A}_{\textbf{x}}&\equiv\mathcal{A}(a,Q_{\textbf{x},i},\pi_a,P_{\textbf{x},i})\,,\\
\mathcal{B}_{\textbf{x}}&\equiv\mathcal{B}(a,Q_{\textbf{x},i},\pi_a,P_{\textbf{x},i})\,.
\Eeq

The lattice Hamilton's equations can be now expressed as
\Beq
\label{eq:HamEqLatt}
\partial_{\tau}a&=\left\{a,H_0^{\rm latt}\right\}_{\rm P}^{\rm latt}\,,\\
\partial_{\tau}\pi_a&=\left\{p_a,H_0^{\rm latt}\right\}_{\rm P}^{\rm latt}\,,\\
\partial_{\tau}Q_{\textbf{x},i}&=\left\{Q_{\textbf{x},i},H_0^{\rm latt}\right\}_{\rm P}^{\rm latt}\,,\\
\partial_{\tau}P_{\textbf{x},i}&=\left\{P_{\textbf{x},i},H_0^{\rm latt}\right\}_{\rm P}^{\rm latt}\,,
\Eeq
where again the first and third lines are the definitions of the lattice conjugate momenta from Eq. \eqref{eq:LattMomenta}, whilst the second and fourth lines are the evolution equations for the scale factor and the lattice fields, respectively. We are again missing the (lattice version of the) Gauss constraint. Just like before, it can be recovered from a residual gauge symmetry of the lattice action, Eq. \eqref{eq:SA0lat}, as we show in the following section.

\subsubsection{Gauss constraint}

Thanks to the links and plaquettes the lattice action given in Eq. \eqref{eq:SA0lat} is invariant under the {\it time-independent} symmetry transformations
\Beq
\varphi_{\textbf{x}}\rightarrow\Omega_{\textbf{x}}\varphi_{\textbf{x}}\,,\quad U_{j,\textbf{x}}\rightarrow\Omega_{\textbf{x}}U_{j,\textbf{x}}\Omega_{\textbf{x}+\textbf{n}_jb}^*\,\quad{\rm where}\quad\Omega_{\textbf{x}}=\exp[-i\alpha_{\textbf{x}}]\,.
\Eeq
In their infinitesimal form ($\alpha_{\textbf{x}}\ll1$),
\Beq
\delta\varphi_{1,\textbf{x}}=\varphi_{2,\textbf{x}}\alpha_{\textbf{x}}\,,\quad\delta\varphi_{2,\textbf{x}}=-\varphi_{1,\textbf{x}}\alpha_{\textbf{x}}\,,\quad\delta A_{j,\textbf{x}}=\frac{\alpha_{\textbf{x}+\textbf{n}_jb}-\alpha_{\textbf{x}}}{b}\,.
\Eeq
The Noether theorem then implies that for arbitrary $\alpha_{\textbf{x}}$, there is a conserved charge
\Beq
J(\tau)&=\sum_{\textbf{x}}^{N^3}\sum_{i=1}^5\frac{\partial L_{0}^{\rm latt}}{\partial Q'_{\textbf{x},i}}\delta Q_{\textbf{x},i}\\
&=\sum_{\textbf{x}}^{N^3}\left[\alpha_{\textbf{x}}(\pi_{1,\textbf{x}}(\tau)\varphi_{2,\textbf{x}}(\tau)-\pi_{2,\textbf{x}}(\tau)\varphi_{1,\textbf{x}}(\tau))+\sum_j^{1,2,3}\pi_{Aj,\textbf{x}}(\tau)\frac{\alpha_{\textbf{x}+\textbf{n}_jb}-\alpha_{\textbf{x}}}{b}\right]\\
&=\text{const}\,.
\Eeq
It reduces to\footnote{We have relabeled the dummy indices inside the last summation in the square brackets, as allowed by the periodic boundary conditions on the lattice.}
\Beq
J(\tau)=\sum_{\textbf{x}}^{N^3}\alpha_{\textbf{x}}\left[\left(\pi_{1,\textbf{x}}(\tau)\varphi_{2,\textbf{x}}(\tau)-\pi_{2,\textbf{x}}(\tau)\varphi_{1,\textbf{x}}(\tau)\right)-\sum_j^{1,2,3}\frac{\pi_{Aj,\textbf{x}}(\tau)-\pi_{Aj,\textbf{x}-\textbf{n}_jb}(\tau) }{b}\right]=\text{const}\,.
\Eeq
Note that $J'(\tau)=0$ holds for arbitrary $\alpha_{\textbf{x}}$. Hence, at each lattice point
\Beq
\label{eq:DGauss}
e^{-2}\mathcal{C}_{\rm G,\textbf{x}}^{\rm latt}(\tau)=\sum_{j}^{1,2,3}\left[\frac{\pi_{Aj,\textbf{x}}(\tau)-\pi_{Aj,\textbf{x}-\textbf{n}_jb}(\tau)}{b}\right]-\left(\pi_{1,\textbf{x}}(\tau)\varphi_{2,\textbf{x}}(\tau)-\pi_{2,\textbf{x}}(\tau)\varphi_{1,\textbf{x}}(\tau)\right)=\text{const}\,.
\Eeq
This is the lattice counterpart to the Gauss constraint given in Eq. \eqref{eq:CGauss}. Thus, the gauge-fixed lattice action, Eq. \eqref{eq:SA0lat}, again enforces $\partial_{\tau}\mathcal{C}_{\rm G,\textbf{x}}^{\rm latt}=0$ and so do its Hamilton's equations. The desired Abelian-Higgs theory with $\mathcal{C}_{\rm G,\textbf{x}}^{\rm latt}(\tau)=0$ is recovered by imposing the Gauss constraint on the initial conditions and then using the lattice Hamilton's equations to evolve the system.

The remaining question is whether there is a way to discretize in time the lattice fields, so that their evolution (according to the corresponding discretized in time lattice Hamilton's equations) still respects the Gauss constraint in Eq. \eqref{eq:DGauss}. We show in the next section, that there is a scheme based on symplectic discretization which does that.

\section{GFiRe implementation}
\label{sec:GFiReImplementation}

As we argue below, we can numerically time-evolve the Abelian-Higgs system prescribed by the lattice action in \eqref{eq:SA0lat} and meet the Gauss constraint \eqref{eq:DGauss} everywhere on the lattice using symplectic integrators. 

In ${\sf GFiRe}$, we use a symplectic prescription to integrate the Hamilton's equations on the lattice (Eqns. \eqref{eq:HamEqLatt}), which we write compactly as
\Beq
\label{eq:HamiltonEquations}
z_j'(\tau)&=\left\{z_j(\tau),H_{0}^{\rm latt}\Big(z_j(\tau)\Big)\right\}_{\rm P}^{\rm latt}\,,
\Eeq
where $z=[a,\varphi_{1,\textbf{x}},\varphi_{2,\textbf{x}},A_{1,\textbf{x}},A_{2,\textbf{x}},A_{3,\textbf{x}},p_a,\pi_{1,\textbf{x}},\pi_{2,\textbf{x}},\pi_{A1,\textbf{x}},\pi_{A2,\textbf{x}},\pi_{A3,\textbf{x}}]^T$, i.e., the integer $j$ index of $z$ runs over the $2\times\left(1+5\times N^3\right)$ phase space variables. In the rest of the section we give the details of the numerical integrator.

\begin{figure}
 \centering
 \includegraphics[width=6.0in]{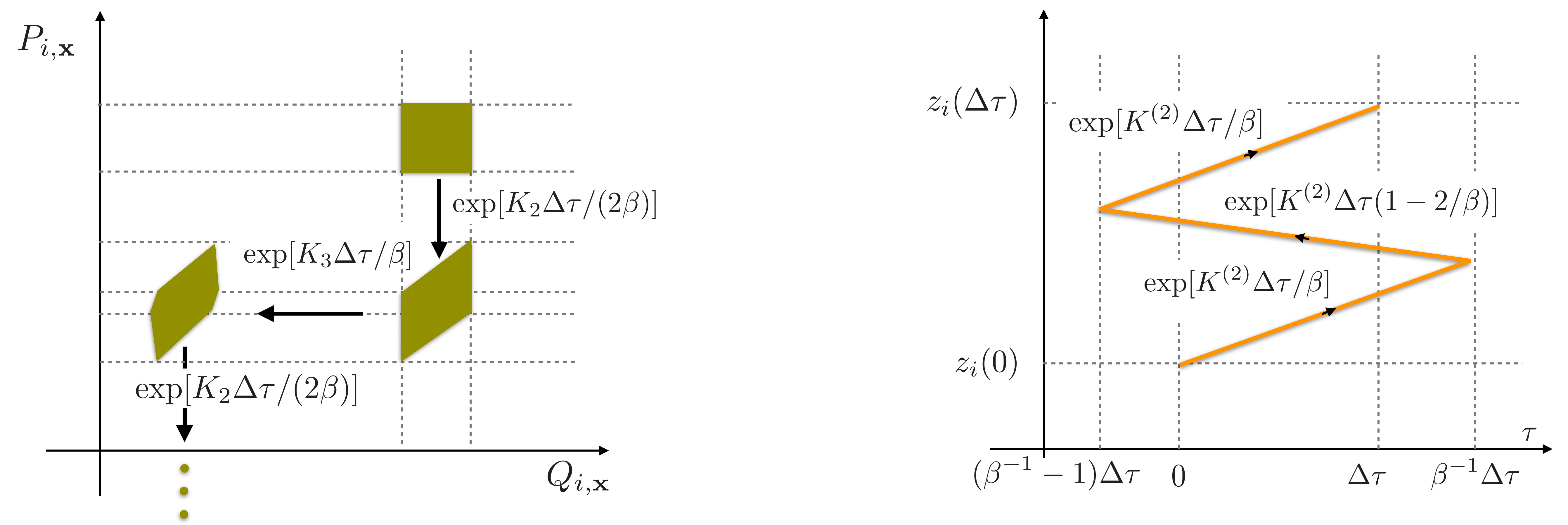}
 \caption{A visual representation our composite symplectic time integrator. The left panel shows the symplectic nature of the integrator, whereas the right panel describes the composition of the building block of the symplectic integrator.}
 \label{fig:SymplecticTimeEvolve}
\end{figure}
\subsection{Symplectic integrators}

We first present the structure of the symplectic integrator used by ${\sf GFiRe}$ in an abstract form. It might be useful to refer to Fig.~\ref{fig:SymplecticTimeEvolve} for a schematic picture of the integrator.

 We need two assumptions about the form of the Hamiltonian. First, we assume that the Hamiltonian splits into three mutually non-commuting terms 
\Beq
\label{eq:HamiltonSpli}
&H=H_1+H_2+H_3\,,\\
&\left\{H_i,H_j\right\}_{{}_\text{P}}\neq0\,.
\Eeq
Second, we also assume that each Hamiltonian term is a function of only one variable from each pair of generalized coordinate and corresponding conjugate momentum. In other words, for a given $j$ we have either $\partial H_i/\partial q_j=0$ or $\partial H_i/\partial p_j=0$ or both (here $p_j$ is the conjugate momentum to the generalised coordinate $q_j$). This assumption allows us to use standard symplectic integrator techniques \cite{Yoshida:1990zz}.

The formal solution to the Hamilton's equations \eqref{eq:HamiltonEquations}, can be written as\footnote{Note that in this formal solution, $H$ is constant in time, since it is evaluated on the solution.}
\Beq
\label{eq:Formal}
z_j(\tau+\Delta\tau)=\exp[\left\{\cdot,H\right\}_{{}_\text{P}} \Delta\tau]z_j(\tau)\,,
\Eeq
for an arbitrary time interval $\Delta \tau$.  The numerical approximation for any finite time interval  $\Delta\tau$ is (see \cite{Yoshida:1990zz}):
\Beq
\label{eq:Integrator}
z_j(\tau+\Delta\tau)=\exp[K^{(k)}\Delta\tau]z_j(\tau)\,,
\Eeq
where $\Delta \tau$ is the time step and $K^{(k)}$ is an operator involving Poisson brackets with the Hamiltonian terms from Eq. \eqref{eq:HamiltonSpli}. The choice of the integrator order, $k$, determines the exact form of the operator. Thanks to our two assumptions, we can use a symplectic method of integration based on operator splitting techniques \cite{Yoshida:1990zz}. The integrator of lowest order, $k=2$, is 
\Beq
\label{eq:K2g}
\exp[K^{(2)}\Delta\tau]&\equiv\exp[K_1\Delta\tau/2]\exp[K_2\Delta\tau/2]\exp[K_3\Delta\tau]\exp[K_2\Delta\tau/2]\exp[K_1\Delta\tau/2]\\
                                       &=\exp[\{\cdot,H\}_{{}_\text{P}}\Delta\tau]+\mathcal{O}(\Delta\tau^3)\,,\\
                                  K_i&=\{\cdot,H_i\}_{{}_\text{P}}\,,
\Eeq
whereas higher-order operators are obtained recursively\footnote{There is a more optimal way of choosing the ``weights'' for higher order composite integrators. Such weights prevent the number of compositions from growing too rapidly as we go to higher and higher integrators (see \cite{Yoshida:1990zz}).} 
\Beq
\label{eq:Kk}
\exp[K^{(k+2)}\Delta\tau]&\equiv\exp[K^{(k)}\Delta\tau/\beta]\exp[K^{(k)}\Delta\tau(1-2\beta^{-1})]\exp[K^{(k)}\Delta\tau/\beta]\\
                                           &=\exp[\{\cdot,H\}_{{}_\text{P}}\Delta\tau]+\mathcal{O}(\Delta\tau^{k+3})\,,\\
                                  \beta&=2-2^{1/(k+1)}\,.
\Eeq

The numerical solution can be made arbitrarily close to the true one as one increases $k$ and/or decreases $\Delta \tau$. Hence, an integrator of arbitrary accuracy has been reduced to the operation of $\exp[K_1\Delta\tau]$, $\exp[K_2\Delta\tau]$ and $\exp[K_3\Delta\tau]$ in some particular combination and taking the appropriate values for the time interval $\Delta\tau$ for each operation. Let's see how this can be implemented for the Abelian-Higgs system.

\subsection{Evolution on the lattice}

The lattice Hamiltonian, $H_{0}^{\rm latt}$, from Eq. \eqref{eq:HamiltonianA0FRWlatt}, splits into three non-commuting terms, just like in Eq. \eqref{eq:HamiltonSpli},
\Beq
\label{eq:HamiltonianA0FRWLatt}
&H_{0}^{\rm latt}=H_1+H_2+H_3\,,\\
 &H_1=-\frac{\pi_a^2}{12\mpl^2N^3}\,,\qquad H_2=\sum_{\textbf{x}}^{N^3}\left[\frac{\pi_{1,\textbf{x}}^2}{2a^2}+\frac{\pi_{2,\textbf{x}}^2}{2a^2}+e^2\sum_j^{1,2,3}\frac{\pi_{Aj,\textbf{x}}^2}{2}\right]\,,\\
&H_3=\sum_{\textbf{x}}^{N^3}\Bigg\{a^4V_{\textbf{x}}+\frac{1}{2e^2b^4}\sum_{i,j}^{1,2,3}\left[1-P_{ij,\textbf{x}}\right]+\frac{a^2}{2b^2}\sum_j^{1,2,3}\bigg[(\varphi_{1,\textbf{x}+\textbf{n}_jb}^2+\varphi_{2,\textbf{x}+\textbf{n}_jb}^2)\\
&\qquad \qquad +(\varphi_{1,\textbf{x}}^2+\varphi_{2,\textbf{x}}^2)-\bigg((\varphi_{1,\textbf{x}}-i\varphi_{2,\textbf{x}})U_{j,\textbf{x}}(\varphi_{1,\textbf{x}+\textbf{n}_jb}+i\varphi_{2,\textbf{x}+\textbf{n}_jb})+c.c.\bigg) \bigg]\Bigg\}\,.
\Eeq
Note that each of the terms is a function either of a generalised coordinate or its conjugate momentum, but never of both consistent with our second assumption in the previous, more formal subsection.

We can now use Eqs. \eqref{eq:Integrator}, \eqref{eq:K2g} and \eqref{eq:Kk} to evolve the complex scalar, 3-vector gauge field, scale factor and their conjugate momenta. To this end we must find the action of each $\exp[K_i\Delta\tau]$ on the lattice phase space variables. In deriving the expressions below, we make an extensive use of the operator identity for our Hamiltonian \eqref{eq:HamiltonianA0FRWlatt}: 
\Beq
\exp[K_i\Delta\tau]z_j=(1+K_i\Delta\tau)z_j\,,
\Eeq
when applied to any $z_j$. Note that this is not an approximation in the context of our Hamiltonian.

After acting with $\exp[K_1\Delta\tau/2]$ on all $z_j$ we get (see Eq. \eqref{eq:Integrator})
\Beq
\label{eq:K1}
a\rightarrow a-\left(\frac{\Delta\tau}{2}\right)\frac{\pi_a}{6\mpl^2N^3}\,,
\Eeq
with the rest of $z_j$ unchanged. Similarly, the action of $\exp[K_2\Delta\tau/2]$ yields
\Beq
\label{eq:K2}
\varphi_{1,\textbf{x}}&\rightarrow\varphi_{1,\textbf{x}}+\left(\frac{\Delta\tau}{2}\right)\frac{\pi_{1,\textbf{x}}}{a^2}\,,\quad\varphi_{2,\textbf{x}}\rightarrow\varphi_{2,\textbf{x}}+\left(\frac{\Delta\tau}{2}\right)\frac{\pi_{2,\textbf{x}}}{a^2}\,,\\
 A_{j,\textbf{x}}&\rightarrow A_{j,\textbf{x}}+\left(\frac{\Delta\tau}{2}\right)e^2\pi_{Aj,\textbf{x}}\,,\quad\pi_a\rightarrow\pi_a+\left(\frac{\Delta\tau}{2}\right)\sum_{\textbf{x}}^{N^3}\left[\frac{\pi_{1,\textbf{x}}^2}{a^3}+\frac{\pi_{2,\textbf{x}}^2}{a^3}\right]\,,
\Eeq
leaving the rest of the generalized coordinates and momenta unchanged. Finally, the action of $\exp[K_3\Delta\tau]$ gives
\Beq
\label{eq:K3}
\pi_{1,\textbf{x}}&\rightarrow\pi_{1,\textbf{x}}+\Delta\tau\Bigg\{-a^4\frac{\partial V_{\textbf{x}}}{\partial\varphi_{1,\textbf{x}}}-\frac{6a^2\varphi_{1,\textbf{x}}}{b^2}+\frac{a^2}{2b^2}\sum_{j}^{1,2,3}\Bigg[U_{j,\textbf{x}}(\varphi_{1,\textbf{x}+\textbf{n}_jb}+i\varphi_{2,\textbf{x}+\textbf{n}_jb})\\
&\qquad\qquad\qquad\qquad\qquad\qquad\qquad\qquad\qquad+(\varphi_{1,\textbf{x}-\textbf{n}_jb}-i\varphi_{2,\textbf{x}-\textbf{n}_jb})U_{j,\textbf{x}-\textbf{n}_jb}+c.c.\Bigg]\Bigg\}\,,\\
\pi_{2,\textbf{x}}&\rightarrow\pi_{2,\textbf{x}}+\Delta\tau\Bigg\{-a^4\frac{\partial V_{\textbf{x}}}{\partial\varphi_{2,\textbf{x}}}-\frac{6a^2\varphi_{2,\textbf{x}}}{b^2}+\frac{a^2}{2b^2}\sum_{j}^{1,2,3}\Bigg[-iU_{j,\textbf{x}}(\varphi_{1,\textbf{x}+\textbf{n}_jb}+i\varphi_{2,\textbf{x}+\textbf{n}_jb})\\
&\qquad\qquad\qquad\qquad\qquad\qquad\qquad\qquad\qquad+i(\varphi_{1,\textbf{x}-\textbf{n}_jb}-i\varphi_{2,\textbf{x}-\textbf{n}_jb})U_{j,\textbf{x}-\textbf{n}_jb}+c.c.\Bigg]\Bigg\}\,,\\
\pi_{Aj,\textbf{x}}&\rightarrow\pi_{Aj,\textbf{x}}+\Delta\tau\Bigg\{\frac{a^2}{2b}\left[i(\varphi_{1,\textbf{x}}-i\varphi_{2,\textbf{x}})U_{j,\textbf{x}}(\varphi_{1,\textbf{x}+\textbf{n}_jb}+i\varphi_{2,\textbf{x}+\textbf{n}_jb})+c.c.\right]\\
&\qquad\qquad\qquad\qquad\qquad\qquad\qquad\qquad\qquad+\frac{1}{2e^2b^3}\sum_l^{1,2,3}\left[iP_{jl,\textbf{x}}+iP_{lj,\textbf{x}-\textbf{n}_lb}+c.c.\right]\Bigg\}\,,\\
\pi_a&\rightarrow\pi_a+\Delta\tau\sum_{\textbf{x}}^{N^3}\Bigg\{-4a^3V_{\textbf{x}}-\frac{a}{b^2}\sum_j^{1,2,3}\bigg[(\varphi_{1,\textbf{x}+\textbf{n}_jb}^2+\varphi_{2,\textbf{x}+\textbf{n}_jb}^2)+(\varphi_{1,\textbf{x}}^2+\varphi_{2,\textbf{x}}^2)\\
&\qquad\qquad\qquad\qquad\qquad\qquad-\bigg((\varphi_{1,\textbf{x}}-i\varphi_{2,\textbf{x}})U_{j,\textbf{x}}(\varphi_{1,\textbf{x}+\textbf{n}_jb}+i\varphi_{2,\textbf{x}+\textbf{n}_jb})+c.c.\bigg) \bigg]\Bigg\}\,,
\Eeq
with no changes in the generalised coordinates. Below, we state one of the main benefits of using the above scheme for time evolution.
\subsection{Preservation of the Gauss constraint}
We verify explicitly that after each individual step, Eq. \eqref{eq:K1} and/or Eq. \eqref{eq:K2} and/or Eq. \eqref{eq:K3}, the lattice Gauss constraint, Eq. \eqref{eq:DGauss}, is respected exactly, $\mathcal{C}_{\rm G,\textbf{x}}^{\rm latt}(\tau)=\text{const}$.\footnote{ It would be of enormous value to prove the result regarding the machine precision preservation of the Gauss constraint by the symplectic time evolution (which we showed for our specific Hamiltonian) more generally. A step in this direction is to note that the Gauss constraint in Eq. \eqref{eq:DGauss} commutes with $H_1$, $H_2$ and $H_3$ separately.} 

\noindent To see this, first note that the action of $\exp[K_1\Delta\tau/2]$ (see Eq. \eqref{eq:K1}) trivially yields $\mathcal{C}_{\rm G,\textbf{x}}^{\rm latt}\rightarrow \mathcal{C}_{\rm G,\textbf{x}}^{\rm latt}$. 
The $\exp[K_2\Delta\tau/2]$ step (see Eq. \eqref{eq:K2}) gives
\Beq
\mathcal{C}_{\rm G,\textbf{x}}^{\rm latt}\rightarrow \mathcal{C}_{\rm G,\textbf{x}}^{\rm latt}-e^2\left[\pi_{1,\textbf{x}}\left(\frac{\Delta\tau}{2}\right)\frac{\pi_{2,\textbf{x}}}{a^2}-\pi_{2,\textbf{x}}\left(\frac{\Delta\tau}{2}\right)\frac{\pi_{1,\textbf{x}}}{a^2}\right]=\mathcal{C}_{\rm G,\textbf{x}}^{\rm latt}\,.
\Eeq
Finally, the $\exp[K_3\Delta\tau]$ (see Eq. \eqref{eq:K3}) operation yields
\Beq
\mathcal{C}_{\rm G,\textbf{x}}^{\rm latt}&\rightarrow \mathcal{C}_{\rm G,\textbf{x}}^{\rm latt}+\frac{e^2}{b}\sum_{j}^{1,2,3}
\Delta\tau\Bigg\{\frac{a^2}{2b}\left[i(\varphi_{1,\textbf{x}}-i\varphi_{2,\textbf{x}})U_{j,\textbf{x}}(\varphi_{1,\textbf{x}+\textbf{n}_jb}+i\varphi_{2,\textbf{x}+\textbf{n}_jb})+c.c.\right]\\
&\qquad\qquad\qquad\qquad\qquad\qquad\qquad\qquad+\frac{1}{2e^2b^3}\sum_l^{1,2,3}\left[iP_{jl,\textbf{x}}+iP_{lj,\textbf{x}-\textbf{n}_lb}+c.c.\right]\Bigg\}\\
&-\frac{e^2}{b}\sum_{j}^{1,2,3}
\Delta\tau\Bigg\{\frac{a^2}{2b}\left[i(\varphi_{1,\textbf{x}-\textbf{n}_jb}-i\varphi_{2,\textbf{x}-\textbf{n}_jb})U_{j,\textbf{x}-\textbf{n}_jb}(\varphi_{1,\textbf{x}}+i\varphi_{2,\textbf{x}})+c.c.\right]\\
&\qquad\qquad\qquad\qquad\qquad\qquad+\frac{1}{2e^2b^3}\sum_l^{1,2,3}\left[iP_{jl,\textbf{x}-\textbf{n}_jb}+iP_{lj,\textbf{x}-\textbf{n}_jb-\textbf{n}_lb}+c.c.\right]\Bigg\}\\
&-e^2\varphi_{2,\textbf{x}}\Delta\tau\Bigg\{-a^4\frac{\partial V_{\textbf{x}}}{\partial\varphi_{1,\textbf{x}}}-\frac{6a^2\varphi_{1,\textbf{x}}}{b^2}+\frac{a^2}{2b^2}\sum_{j}^{1,2,3}\Bigg[U_{j,\textbf{x}}(\varphi_{1,\textbf{x}+\textbf{n}_jb}+i\varphi_{2,\textbf{x}+\textbf{n}_jb})\\
&\qquad\qquad\qquad\qquad\qquad\quad\qquad\qquad+(\varphi_{1,\textbf{x}-\textbf{n}_jb}-i\varphi_{2,\textbf{x}-\textbf{n}_jb})U_{j,\textbf{x}-\textbf{n}_jb}+c.c.\Bigg]\Bigg\}\\
&+e^2\varphi_{1,\textbf{x}}\Delta\tau\Bigg\{-a^4\frac{\partial V_{\textbf{x}}}{\partial\varphi_{2,\textbf{x}}}-\frac{6a^2\varphi_{2,\textbf{x}}}{b^2}+\frac{a^2}{2b^2}\sum_{j}^{1,2,3}\Bigg[-iU_{j,\textbf{x}}(\varphi_{1,\textbf{x}+\textbf{n}_jb}+i\varphi_{2,\textbf{x}+\textbf{n}_jb})\\
&\qquad\qquad\qquad\qquad\qquad\qquad\quad\qquad+i(\varphi_{1,\textbf{x}-\textbf{n}_jb}-i\varphi_{2,\textbf{x}-\textbf{n}_jb})U_{j,\textbf{x}-\textbf{n}_jb}+c.c.\Bigg]\Bigg\}\\
&=\mathcal{C}_{\rm G,\textbf{x}}^{\rm latt}\,,
\Eeq
where the $\partial V$ terms cancel due to the $U(1)$ symmetry of the potential, the links, $U_j$, terms in the first two big brackets cancel with the links terms in the other two big brackets, whereas the sums of the Plaquette terms vanish by virtue of the identity $\sum_{j,l}^{1,2,3}\Big[iP_{jl,\textbf{x}}+c.c.\Big]=0$. (also see Eq. \eqref{eq:LinksPlaquettes}).

Another, perhaps more impressive way of stating our result is that when we evaluate $\mathcal{C}^{\rm latt}_G$ at $\tau+\Delta\tau$ by using the above expressions for time evolved $z_j$, all terms proportional to $\Delta\tau^n$ for arbitrary $n$ vanish. Thus, regardless of the order of the time integrator, {\it k}, the Gauss constraint is always satisfied to machine precision at each lattice site.\footnote{We emphasize that this holds only for Abelian fields. When the same analysis is repeated for a Higgs doublet coupled to a non-Abelian $SU(2)$ gauge field, the lattice Gauss constraint is violated by the symplectic integrator.} As we showed above, for physical consistency, initially we must set the constant on the right hand side in the Gauss constraint equation, Eq. \eqref{eq:DGauss}, to zero everywhere on the lattice. The symplectic evolution now guarantees that it remains zero at later times. 

\subsection{Approximate post-inflationary initial conditions}
\label{sec:InitialConditions}

\begin{figure}
 \centering
 \includegraphics[width=5.7in]{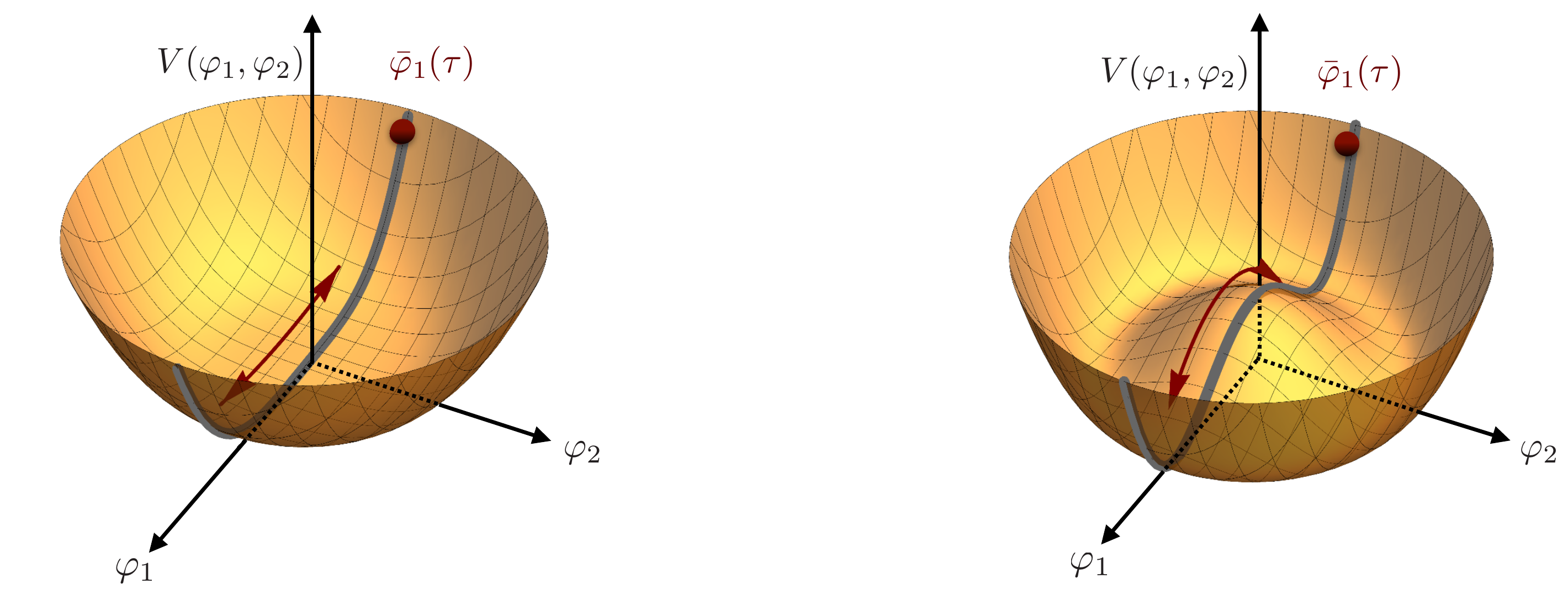}
 \caption{The complex scalar potential profile, see Eq. \eqref{eq:Vinflaton}, for two separate cases $v=0$ (left panel) and $v\ne 0$ (right panel). After inflation ends, the field is primarily rolling down along $\varphi_1$-axis, followed by oscillations about the minimum, which can lead to resonant particle production in the gauge fields as well as the $\vp$ field. Note that in both cases, we begin with $\bar{\vp}_1(\tau_{\rm in})\gg v$.}
 \label{fig:PotBowl}
\end{figure}

Within the inflationary paradigm, the (almost) homogeneous inflaton field dominates the Universe during and shortly after the end of inflation. One can assume that the inflaton is the real part of the complex salar (i.e., slow-roll inflation is realized along the real axis in the complex field space, towards the minimum of the scalar field potential, see Fig. \ref{fig:PotBowl}). At the beginning of our simulations of preheating, the initial variables on the lattice are
\Beq
\label{eq:InitBckgrnd}
&\varphi_{1,\textbf{x}}=\bar{\varphi}_{1}+\delta\varphi_{1,\textbf{x}}\,,\quad\varphi_{2,\textbf{x}}=\delta\varphi_{2,\textbf{x}}\,,\quad \pi_{1,\textbf{x}}=\bar{\pi}_{1}+\delta\pi_{1,\textbf{x}}\,,\quad\pi_{2,\textbf{x}}=\delta\pi_{2,\textbf{x}}\,,\\
& A_{j,\textbf{x}}=\delta A_{j,\textbf{x}}\,,\quad \pi_{Aj,\textbf{x}}=\delta \pi_{Aj,\textbf{x}}\,,\quad a=1\,,\quad\pi_a=-2\sqrt{3}\mpl N^3\sqrt{\frac{H_2+H_3}{N^3}}\,,
\Eeq
where the choice for the initial value of $a$ is conventional and the expression for $\pi_a$ is simply the Friedmann equation. At later times, we use this expression to check the `energy conservation' of our integrators. We observe violations that scale correctly with the time step -- $\mathcal{O}(\Delta\tau^{k})$ for {\it k}-th order integrators. The $\bar{\varphi}_{1}$ and $\bar{\pi}_{1}$ variables are determined from the homogeneous inflationary dynamics, whereas the $\delta$-field perturbations have a power spectrum set by the quantum vacuum fluctuations.

The initial fluctuations for each lattice field can be expanded in terms of Fourier modes.\footnote{The lattice Fourier conventions we use are
\Beq
f_{i,\textbf{x}}=\sum_{\textbf{k}}e^{i\textbf{k}\cdot\textbf{x}}\tilde{f}_{i,\textbf{k}}\,,\quad \tilde{f}_{i,\textbf{k}}=\frac{1}{N^3}\sum_{\textbf{x}}e^{-i\textbf{k}\cdot\textbf{x}}f_{i,\textbf{x}}\,,
\Eeq
implying $\tilde{f}_{i,\textbf{k}}=\tilde{f}_{i,-\textbf{k}}^*$ if $f_{i,\textbf{x}}$ is real.} These can be then written in terms of products of mode functions and `stochastic' complex numbers. The stochasticity in the initial field perturbations is what allows the classical simulations to approximately capture aspects of the quantum uncertainty in the actual vacuum fluctuations. We also must satisfy the Gauss constraint, Eq. \eqref{eq:DGauss}, with the constant term 
\Beq
\label{eq:InitFldsLat}
\delta \varphi_{1,\textbf{x}}&=\frac{(2\pi)^{3/2}}{L^{3/2}_{\text{lat}}}\sum_{\textbf{k}} e^{i\textbf{k}\cdot\textbf{x}} \delta \varphi_{1,\textbf{k}}=\frac{(2\pi)^{3/2}}{L^{3/2}_{\text{lat}}}\sum_{\textbf{k}} e^{i\textbf{k}\cdot\textbf{x}}\Big[a_{\textbf{k}}^1 u_{k}^1+a_{-\textbf{k}}^{1*}u_{ k}^{1*}\Big]\,,\\
\delta \pi_{1,\textbf{x}}&=\frac{(2\pi)^{3/2}}{L^{3/2}_{\text{lat}}}\sum_{\textbf{k}} e^{i\textbf{k}\cdot\textbf{x}} \delta \pi_{1,\textbf{k}}=\frac{(2\pi)^{3/2}}{L^{3/2}_{\text{lat}}}\sum_{\textbf{k}} e^{i\textbf{k}\cdot\textbf{x}}\Big[a_{\textbf{k}}^1 {u_{k}^1}'+a_{-\textbf{k}}^{1*}{u_{ k}^{1*}}'\Big]\,,\\
\delta \varphi_{2,\textbf{x}}&=\frac{(2\pi)^{3/2}}{L^{3/2}_{\text{lat}}}\sum_{\textbf{k}} e^{i\textbf{k}\cdot\textbf{x}} \delta \varphi_{2,\textbf{k}}=\frac{(2\pi)^{3/2}}{L^{3/2}_{\text{lat}}}\sum_{\textbf{k}} e^{i\textbf{k}\cdot\textbf{x}}\Big[a_{\textbf{k}}^2 u_{k}^2+a_{-\textbf{k}}^{2*}u_{ k}^{2*}\Big]\,,\\
\delta \pi_{2,\textbf{x}}&=\delta \varphi_{2,\textbf{x}}\frac{\bar{\pi}_{1}+\delta \pi_{1,\textbf{x}}}{\bar{\varphi}_{1}+\delta \varphi_{1,\textbf{x}}}\,,\\
A_{j,\textbf{x}}&=\frac{(2\pi)^{3/2}}{L^{3/2}_{\text{lat}}}\sum_{\textbf{k}} e^{i\textbf{k}\cdot\textbf{x}} A_{j\textbf{k}}=\frac{(2\pi)^{3/2}}{L^{3/2}_{\text{lat}}}\sum_{\textbf{k}} e^{i\textbf{k}\cdot\textbf{x}}\Big[a_{\textbf{k}}^{Aj} u_{k}^{Aj}+\left(a_{-\textbf{k}}^{Aj}u_{k}^{Aj}\right)^*\Big]\,,\\
\pi_{Aj,\textbf{x}}&=0\,.
\Eeq
The random (`stochastic') complex numbers provide the classical counterpart of the quantum creation and annihilation operators, and take the values
\Beq
\label{eq:stochasticInitNumbers}
\left|a_{\textbf{k}}^I\right|=\sqrt{-\ln\left(X_{\textbf{k}}^I\right)/2}\,,\quad \text{arg}\left(a_{\textbf{k}}^I\right)=2\pi Y_{\textbf{k}}^I\,.
\Eeq
Here $X_{\textbf{k}}^I$ is a uniform deviate on $(0,1)$ and $Y_{\textbf{k}}^I$ is a uniform deviate on $[0,1)$. The mode functions are given by the flat spacetime expressions 
\Beq
u^1_{k}&=\frac{\exp\left(-i\sqrt{k^2+(\partial^2/\partial\bar{\varphi}_1^2)V}\tau\right)}{\sqrt{2}\left(k^2+(\partial^2/\partial\bar{\varphi}_1^2)V\right)^{1/4}}\,,\\
u^2_{k}&=\frac{\exp\left(-i\sqrt{k^2+(\partial^2/\partial\bar{\varphi}_2^2)V}\tau\right)}{\sqrt{2}\left(k^2+(\partial^2/\partial\bar{\varphi}_2^2)V\right)^{1/4}}\,,\\
u^{Aj}_{k}&=\frac{\exp\left(-i\sqrt{k^2+k_{\!_C}^2}\tau\right)}{\sqrt{2}(k^2+k_{\!_C}^2)^{1/4}}\,,
\Eeq
which approximate well the modes of interest, i.e., subhorizon modes, $k\gg\mathcal{H}$, at the end of inflation, regardless of the coupling strength, parametrised by the comoving Compton wavenumber, $ k_{\!_C}\equiv e\bar{\varphi}_1$, for the gauge fields. Note that the normalization factor of $(2\pi)^{3/2}/L^{3/2}_{\text{lat}}$ in Eq. \eqref{eq:InitFldsLat} is needed to make the (initial) two-point functions of field perturbations, averaged over the lattice volume, independent of the comoving box size, $L_{\text{lat}}=Nb$, and equal to the continuous ones, see for example, \cite{Felder:2000hq}. 

\section{Numerical studies of the scalar electrodynamics}
\label{sec:NumStudies}

To test our algorithm, we study the non-linear dynamics in models with the scalar field potential (see Fig.~\ref{fig:PotBowl})
\Beq
\label{eq:Vinflaton}
V=\frac{\lambda}{4}\left(\varphi_1^2+\varphi_2^2-v^2\right)^2\,.
\Eeq
For concreteness, we assume $\varphi_1$ to play the role of the inflaton and set $\lambda=9\times10^{-14}$. The initial conditions at the start of the simulations (and the end of slow-roll inflation) are of the form given in Section \ref{sec:InitialConditions}, with
\Beq
\label{eq:Phi0v}
\bar{\varphi}_1=\phi_0=1.71\mpl\gg v\,.
\Eeq
Note that this setup corresponds to the well-known quartic inflationary scenario \cite{Linde_1984,Linde:2005ht}. As such, this particular shape of the potential is in conflict with CMB observations \cite{Akrami:2018odb}. Nevertheless, since the main focus of this section is on the post-inflationary dynamics, we content ourselves with interpreting Eq. \eqref{eq:Vinflaton} as the effective form of the inflaton potential after inflation, remaining agnostic about the details of $V$ during inflation. For simplicity, we also set $\bar{\pi}_1=0$ initially.

Shortly after the end of inflation $\bar{\varphi}_1(\tau)$ begins to oscillate. Its oscillations can amplify exponentially fast the initial field fluctuations, $\delta\varphi_1$, $\delta\varphi_2$ and $\delta\boldsymbol{\rm{A}}$. The phenomenon can be understood in terms of parametric resonance \cite{Lozanov:2016pac} and is more generally known in the context of inflation as preheating \cite{Kofman:1997yn}. The field fluctuations can grow very quickly, until mode-mode couplings become non-negligible. This marks the end of the linear stage of preheating \cite{Amin:2014eta} and the onset of backreaction. The ensuing non-linear dynamics can be quite rich and is strongly dependent on $v$. We now consider two different scenarios -- $v=0$ and $v\neq0$, separately and study the linear and non-linear regimes with our numerical prescription.

For the results below, we typically rely on a $N^3=512^3$ lattice. We use a fourth order symplectic integrator ($k=2$ in the notation of Eq. \eqref{eq:Kk}) with a time step $\Delta\tau = 0.028(\sqrt{\lambda}\phi_0)^{-1}$ and comoving edge length $L_{\rm latt}=60 (\sqrt{\lambda}\phi_0)^{-1}$. The comoving lattice spacing $\Delta x = L_{\rm latt}/N$. Note that we always have $\Delta\tau <\Delta x$. By the end of the simulations $a(\tau_f)\Delta x< 1/(\sqrt{\lambda}v)$ where $\sqrt{\lambda}v$ is the mass scale in the valley of the potential in the $v\ne 0$ case. In the $v=0$ case, the effective inverse mass scale case grows with the scale factor, hence we always resolve the relevant spatial and temporal scales at late times, if we resolve them initially. 

\begin{figure}
 \centering
 \includegraphics[width=6.0in]{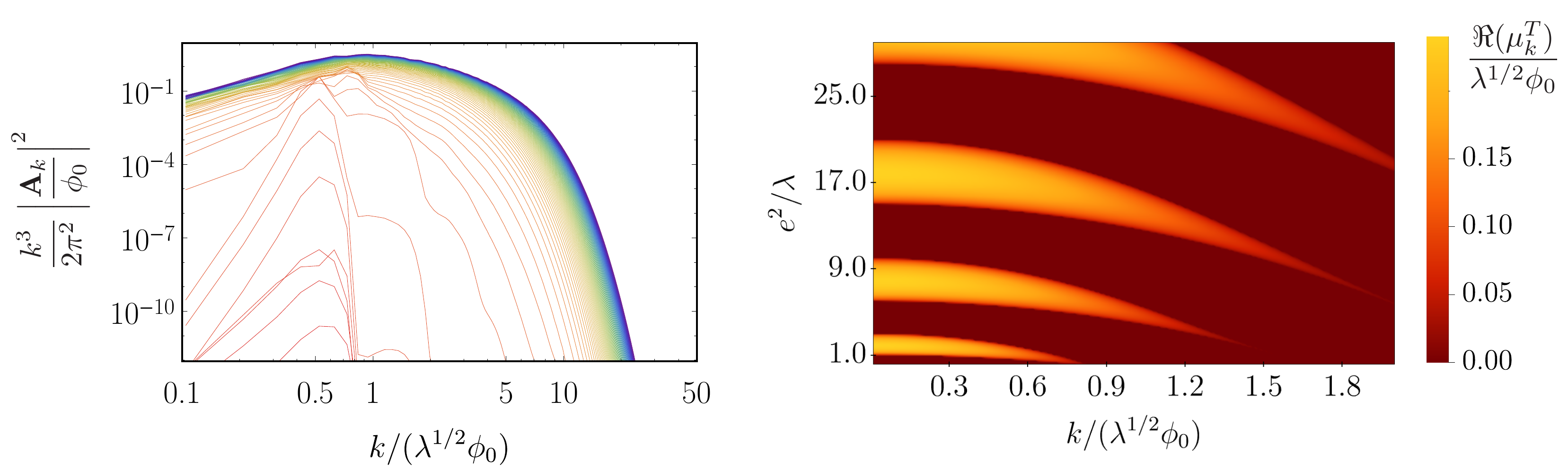}
 \caption{{\it Left panel}: The evolution of the gauge field power spectrum for $v=0$, extracted from our lattice simulations. The different colors correspond to different times, going from red to blue. The initial rapid growth of a broad range of comoving momentum, $k$, modes (dark red curves) is a manifestation of resonant particle production. It is eventually shut off by backreaction of the gauge field on the Higgs condensate. At late times the power distribution becomes time-independent. {\it Right panel}: The Floquet chart corresponding to the instability in the transverse components of the gauge fields. }
 \label{fig:pspNoVevA}
\end{figure}

\subsection{Nonlinear dynamics in the $v= 0$ case}
\label{sec:NoVevPreheating}

We begin with $v=0$, for which the scalar field potential profile takes the form of a quartic bowl, see Fig. \ref{fig:PotBowl}. At the start of the simulations, the background scalar field oscillates along the real axis in the complex $\varphi$ plane, as shown in the left panel of the figure. Depending on the ratio $e^2/\lambda$, these oscillations can lead to the exponential amplification of the complex scalar and/or the gauge field fluctuations. We set
\Beq
\label{eq:elambda}
e=\sqrt{\lambda}\,.
\Eeq
This choice is rather interesting, since it entails significant gauge field, $\delta\boldsymbol{\rm A}$, resonant particle production, but virtually no Higgs particle production\footnote{Note that naively, one would expect also substantial $\delta\varphi_2$ production. Indeed, if one ignores the gauge field fluctuations, in the oscillating $\bar{\varphi}_1(\tau)$ background $\delta\varphi_2$ obeys the Lame equation with a resonant parameter $q=1$, for which we have broad parametric resonance and significant $\delta\varphi_2$ production, see, e.g., \cite{Frolov:2010sz}. However, the linear coupling with the longitudinal gauge field mode modifies the equation of motion of $\delta\varphi_2$ and shuts off the broad resonance for certain choices of $e^2/\lambda$, such as the one in Eq. \eqref{eq:elambda}. For more details, see \cite{Lozanov:2016pac}.}${}^,$\footnote{Irrespective of the parameters, there is always a narrow resonance instability in $\delta\varphi_1$, see, e.g., \cite{PhysRevLett.77.219,Lozanov:2016hid,Lozanov:2017hjm,Lozanov:2019ylm}. For our choice of parameters, Eq. \eqref{eq:elambda}, it is too slow and unimportant when compared to the gauge field instabilities.} \cite{Lozanov:2016pac}. Our lattice simulations capture this subtle effect. 
\begin{figure}[t]
\centering
   \includegraphics[width=3.0in]{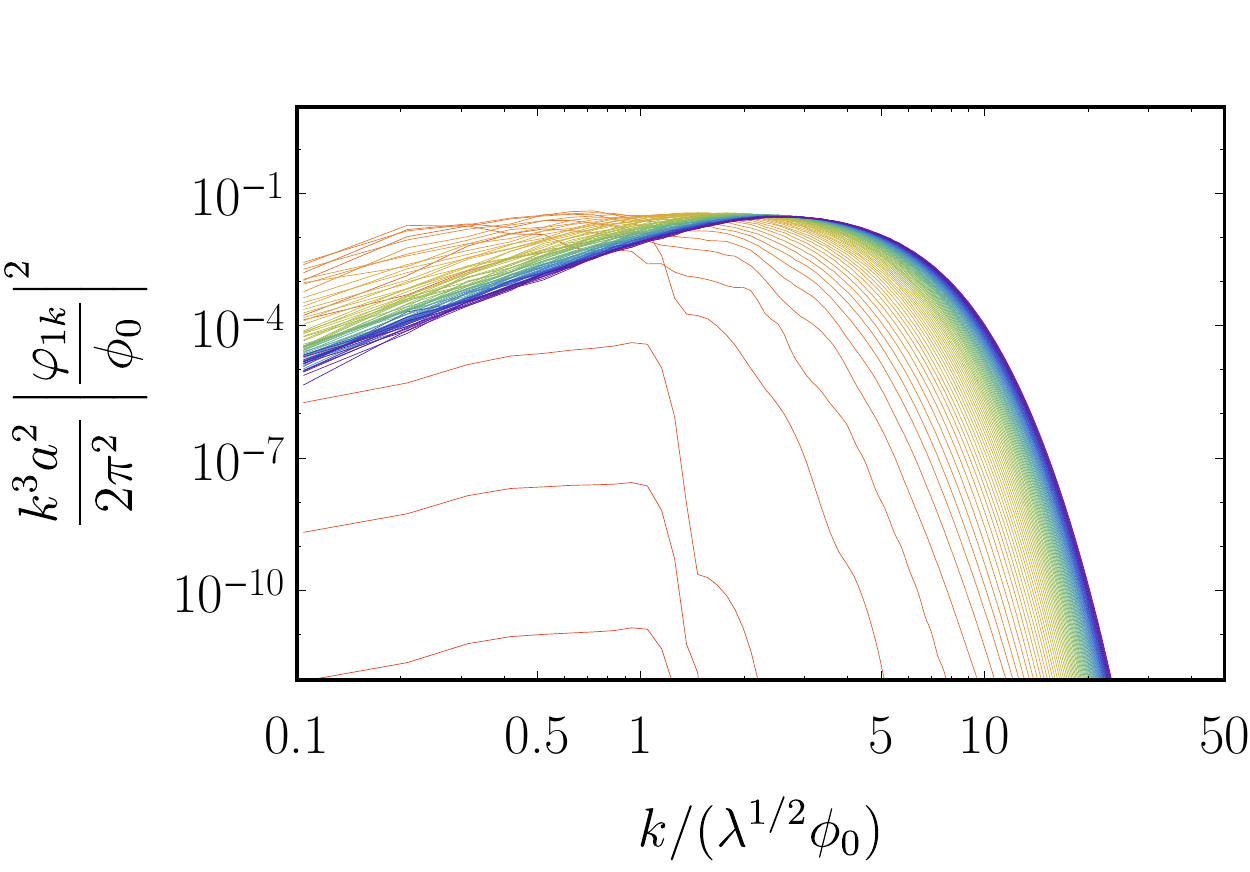}
   \includegraphics[width=3.0in]{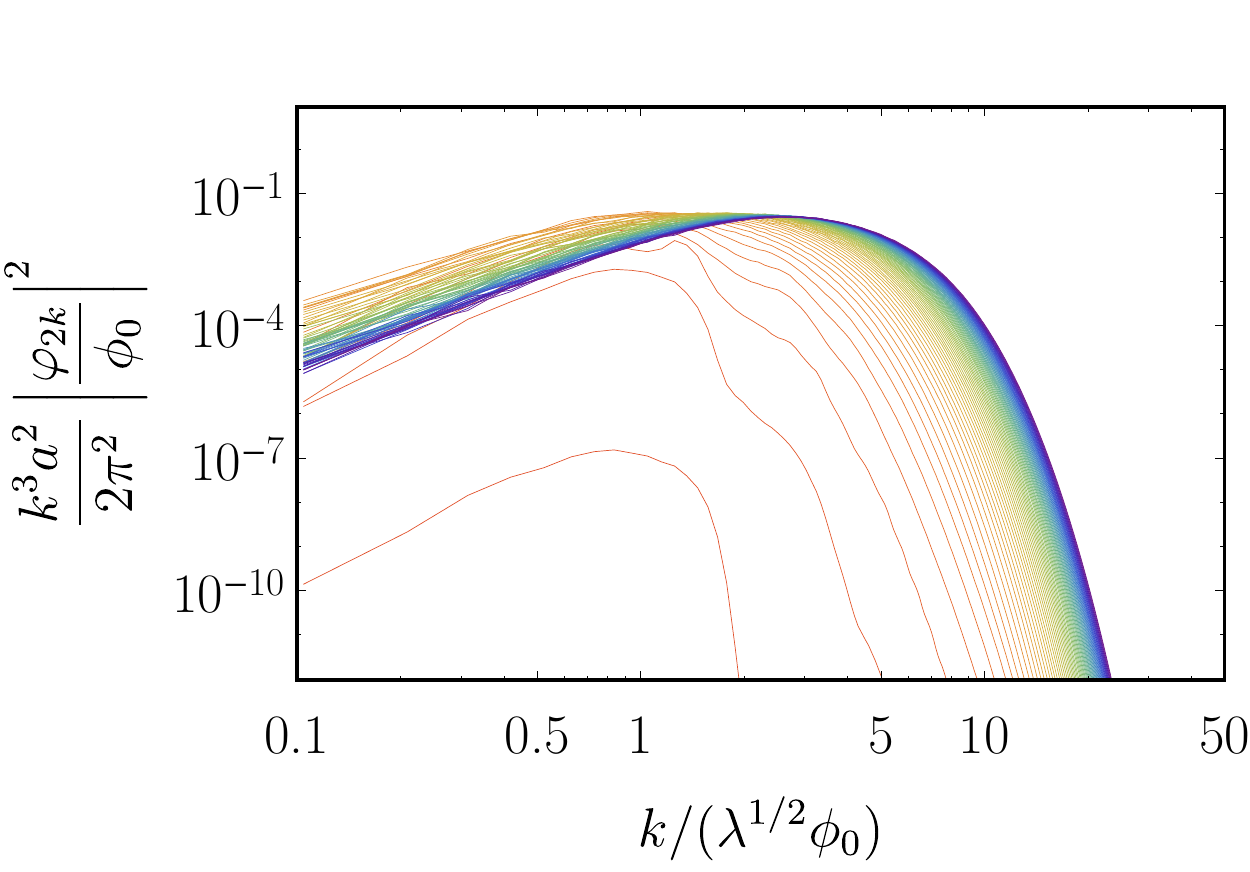}
   \caption{Same as Fig. \ref{fig:pspNoVevA}, but for the real (left panel) and imaginary (right panel) components of the complex field $\vp$. The growth of a broad range of comoving modes depicted by the light red curves is delayed with respect to the growth in the gauge field power spectrum from Fig. \ref{fig:pspNoVevA}. It begins only around the time of backreaction of the gauge field on the $\vp$ condensate. The reason why there is no resonant instability in the real components of $\vp$ (and only in the gauge field) is our choice of parameters, $e^2=\lambda$. For more details see the main text and Ref. \cite{Lozanov:2016pac}. Following backreaction, the power in both components settles to a constant distribution.}
   \label{fig:pspNoVevPhi}
\end{figure}
\\ \\
\noindent{\it Field Power Spectra}: In Fig. \ref{fig:pspNoVevA}, we plot the evolution of the gauge field power spectrum. Soon after the simulations commence, there is a rapid excitation of a broad range of $\delta\boldsymbol{\rm A}$ comoving wavenumbers, as depicted by the dark red curves in the figure. The power spectrum eventually stops growing, when backreaction kicks in. Later on, non-linear effects such as rescattering excite even a wider range of comoving modes and drive the gauge field power spectrum into a long-lived single-broad-peak configuration. 

\begin{figure}[t]
\centering
\includegraphics[width=2.8in]{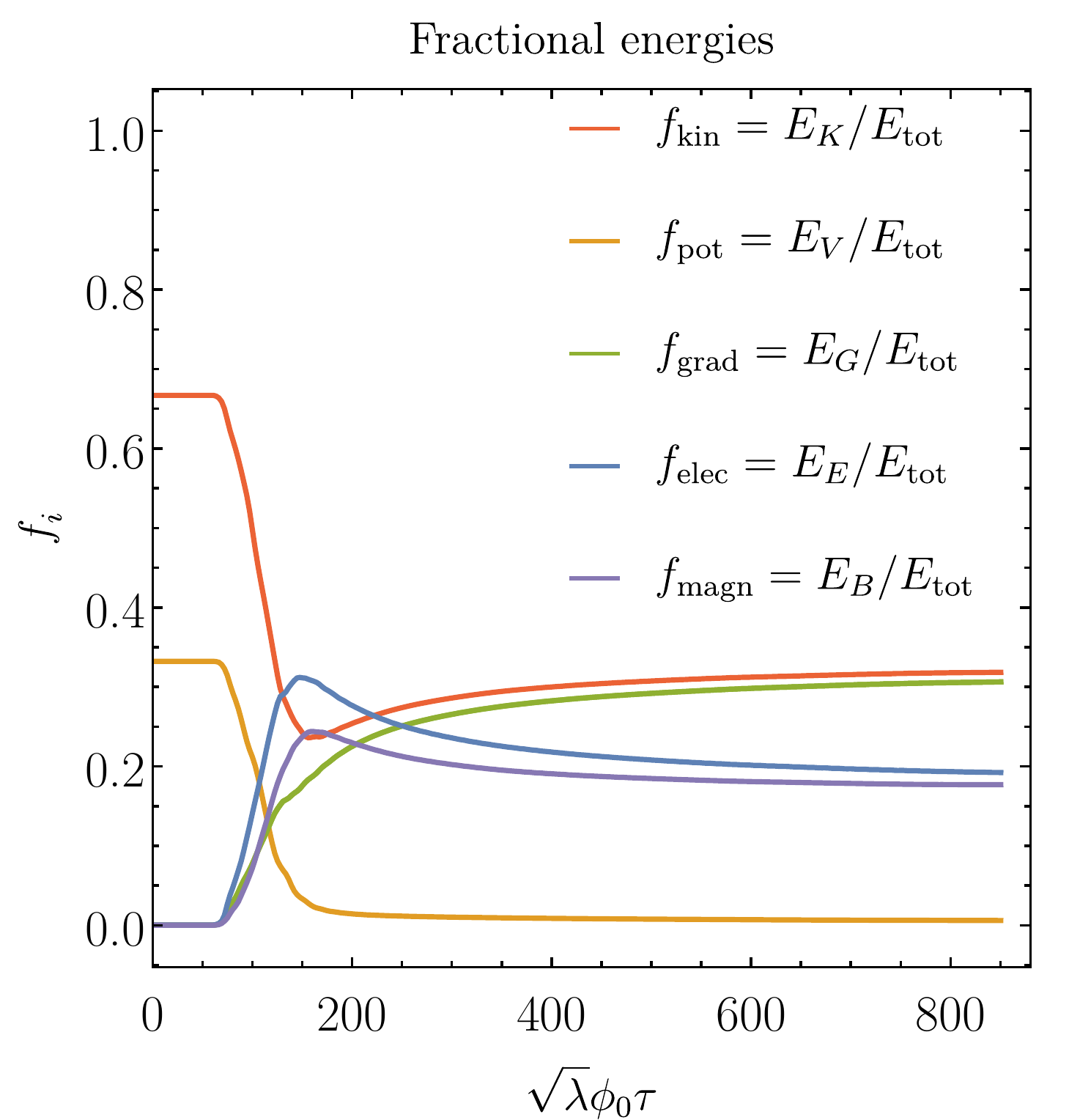}
\includegraphics[width=2.8in]{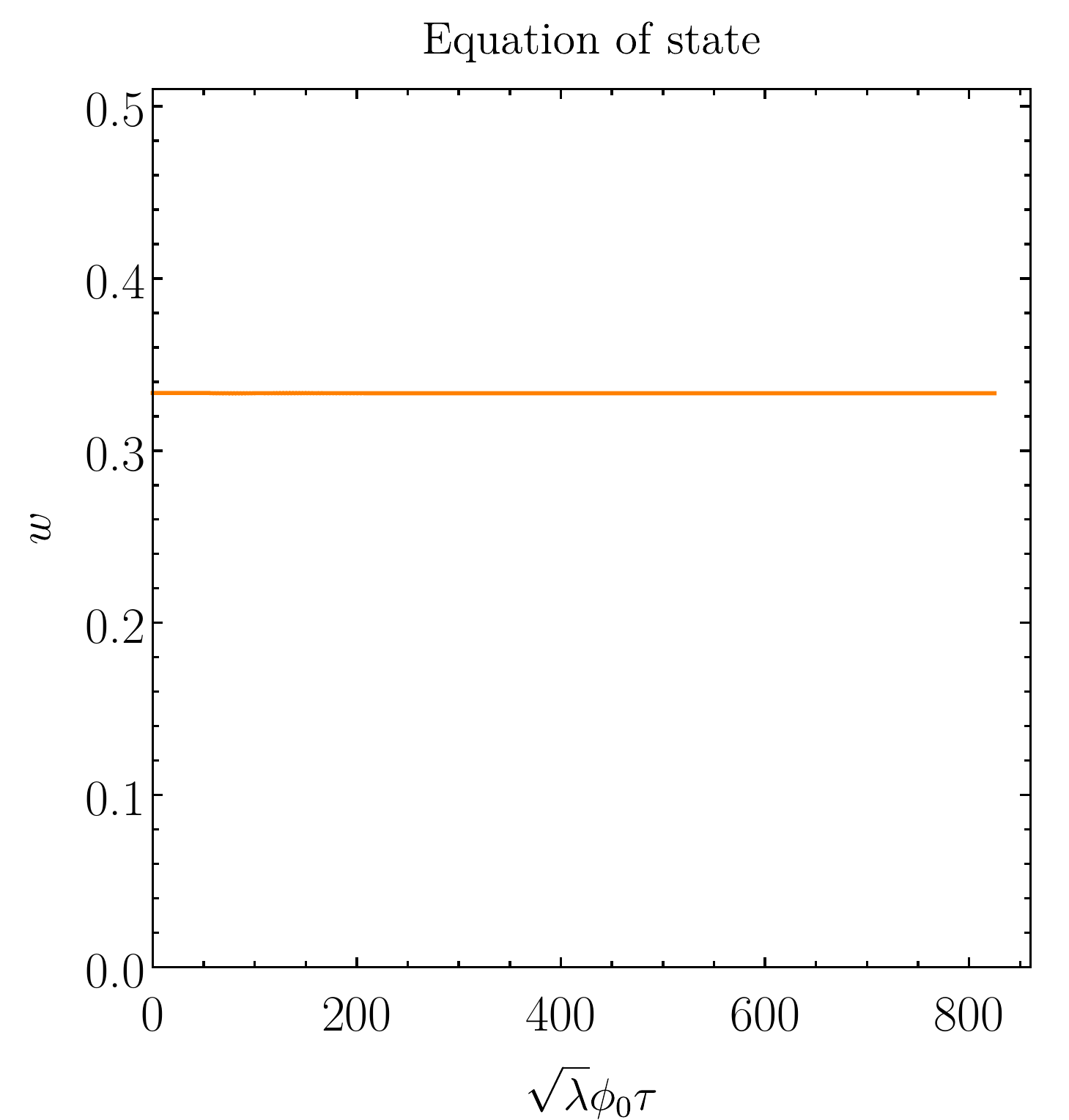}
\caption{The evolution of the normalized energies (left panel), see Eq. \eqref{eq:fisDefn}, and the equation of state (right panel), see Eq. \eqref{eq:EoSDefn}, for the case of $v=0$. All quantities are rapidly oscillating, so the curves are the smoothed time-averages over several oscillations. In the left panel we clearly see the initial oscillatory phase, during which all of the energy is stored in the oscillating $\bar{\varphi}_1$ background in the form of kinetic and potential energy. After backreaction by the gauge fields aournd $\tau\sim70 (\sqrt{\lambda}\phi_0)^{-1}$, energy quickly gets redistributed across all components and the curves approach horizontal asymptotics. At late times, both the Higgs and the gauge field behave as massless radiation, which is reflected by the equation of state on the right.}
\label{fig:NoVevEnergyFraction}
\end{figure}

\begin{figure}[t]
\centering
\includegraphics[width=5.8in]{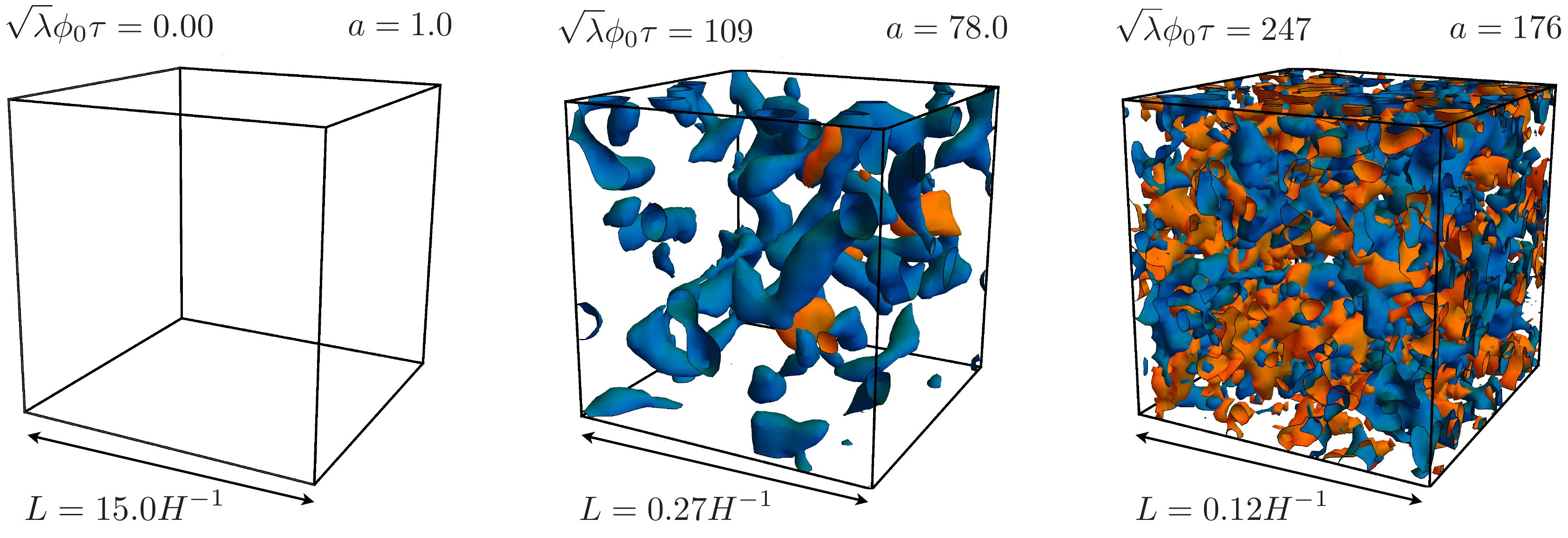}
\caption{The evolution of the normalized electric (blue) and magnetic (orange) field energy densities for the case where $v=0$ (no symmetry breaking). The contours are drawn at $f_i=\rho_i/\rho_{\rm tot}=0.4$ where $i=E,B$. The middle panel is close to the time when backreaction begins. The rightmost panel is at late times after the $\vp$ condensate has fragmented. Note that this figure is produced from half the box compared to the rest of the text.}
\label{fig:EBSlices}
\end{figure}

\begin{figure}[t]
 \centering
 \includegraphics[width=2.8in]{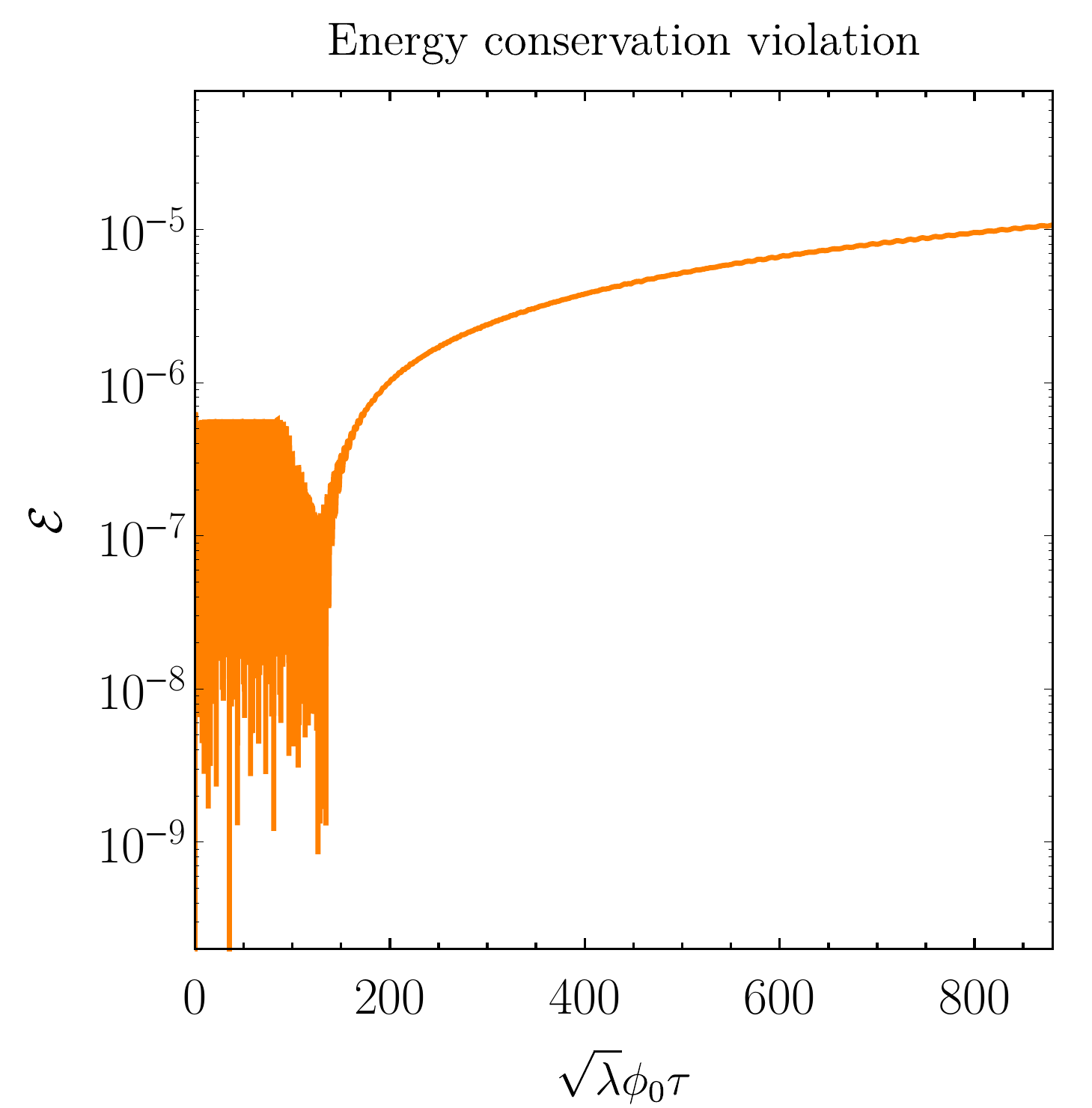}
 \includegraphics[width=2.8in]{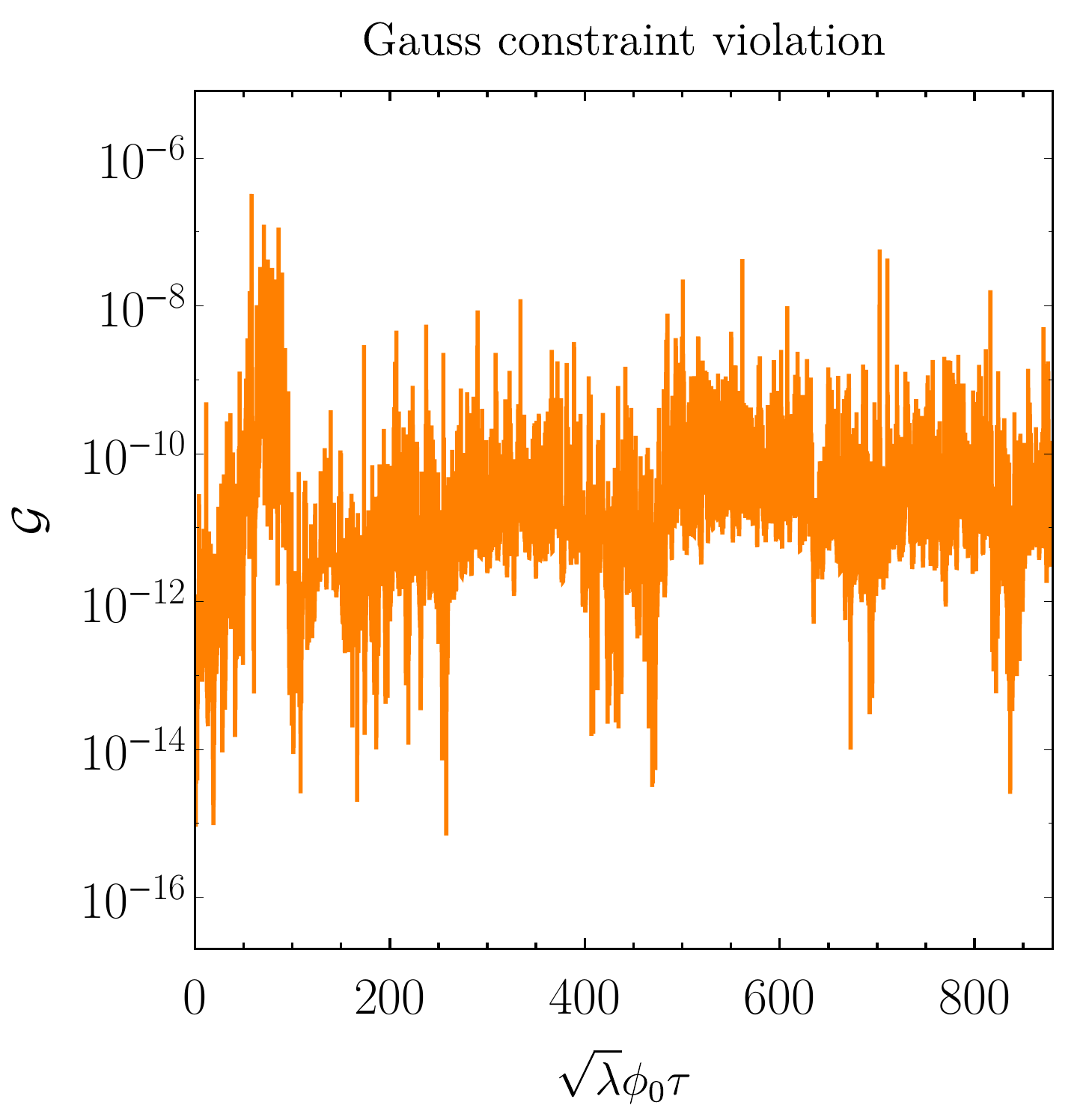}
 \caption{The evolution of the violations of energy conservation (left panel), see Eq. \eqref{eq:CurlyE}, and the Gauss constraint (right panel), see Eq. \eqref{eq:CurlyG} for the case where $v=0$. The energy violation is very stable which is a generic property of symplectic integrators. The Gauss constraint violation starts growing from being close to machine precision initially due to finite differencing noise during the initial homogeneous oscillatory stage. As the Higgs fragments, the violation settles down to a constant small value. Note that we have defined $\mathcal{E}$ and $\mathcal{G}$ to be positive definite.}
 \label{fig:NoVevConserv}
\end{figure}

Unlike the gauge field power spectrum, the power spectra of the two scalar field components, shown in Fig. \ref{fig:pspNoVevPhi}, do not feature the early rapid growth, as expected for our parameter choice, Eq. \eqref{eq:elambda}. A broad range of comoving $\vp$ modes starts getting excited only around the time of backreaction of $\delta\boldsymbol{\rm A}$ on $\bar{\varphi}_1$, as depicted by the light red and orange curves in Fig. \ref{fig:pspNoVevPhi}. At late times, mode-mode couplings again drive the $\vp$ power spectra into a long-lived broad single-peaked configuration.
\\ \\
\noindent{\it Energy fraction and equation of state evolution}: The two-stage evolution (an initial $\bar{\varphi}_1$ oscillatory stage with $\delta\boldsymbol{\rm A}$ growing, followed by a long-lived steady-state nonlinear stage) can be also observed in the evolution of the fractional energies
\Beq
\label{eq:fisDefn}
f_i&\equiv\frac{E_i}{E_{\rm tot}}\,,\qquad E_i=\int d^3\boldsymbol{\rm x} \,\rho_i(\boldsymbol{\rm x})\,,\qquad E_{\rm tot}=\sum_i E_i\,,\\
\rho_K&=\frac{\varphi_1'^2}{2a^2}+\frac{\varphi_2'^2}{2a^2}\,,\qquad \rho_V=V\,,\qquad\rho_{\rm grad}=\frac{|\mathcal{D}_i\varphi|^2}{a^2}\,,\\
\rho_{\rm elec}&=\frac{F_{0j}F_{0j}}{2e^2a^4}\,,\qquad\rho_{\rm magn}=\frac{F_{ij}F_{ij}}{4e^2a^4}\,,
\Eeq
given in the left panel in Fig. \ref{fig:NoVevEnergyFraction}. For $\tau$ between $0$ and $\sim 70(\sqrt{\lambda}\phi_0)^{-1}$, most of the total energy, $E_{\rm tot}$ is stored in the oscillating $\bar{\varphi}_1$, in the form of kinetic and potential energy. Afterwards, as non-linear effects become important the energy gets quickly redistributed across all components. At around $\tau\sim 200(\sqrt{\lambda}\phi_0)^{-1}$, the system enters the long-lived non-linear stage, characterized by a steady energy equipartition \cite{Figueroa:2015rqa}. Thereafter, the $\vp$ self-interaction potential energy vanishes, $f_{\rm pot}\approx 0$, whereas $f_{\rm kin}\approx f_{\rm grad}$ and $f_{\rm elec}\approx f_{\rm magn}$.
Our numerical algorithm also allowed us, for the first time, to compute the self-consistent expansion of the FRW background sourced by an inhomogeneous scalar electrodynamics system. The evolution of the mean equation of state, $w$,
\Beq
\label{eq:EoSDefn}
w=\frac{p}{\rho}=f_K-f_V+\frac{1}{3}(f_{\rm elec}+f_{\rm magn}-f_{\rm grad})\,,
\Eeq
is shown in the right panel in Fig. \ref{fig:NoVevEnergyFraction}. We find that $w=1/3$ throughout, which is to be expected. During the initial oscillatory stage, we have a single homogeneous oscillating scalar field with a quartic self-interaction, which implies the well-known result of $1/3$ for the mean equation of state \cite{Turner:1983he} (in our notation, the only non-zero $f_i$s are $f_K=2f_V=2/3$). Later on, since the $\vp$ self-interaction potential energy vanishes, the real and imaginary parts of $\vp$, as well as the components of the gauge fields behave as massless radiation, again implying a radiation-like equation of state (in our notation in the radiation limit, $f_K+f_{\rm elec}\approx1/2$, since the magnetic and electromagnetic components are approximately equal, as well as the Higgs kinetic and gradient energies).
\\ \\
\noindent {\it Lattice snapshots}: Individual snapshots of the field configurations and their energy densities on the lattice at any given time reveal a rich spatial structure in the fields at both the linear and nonlinear stages. In Fig.~\ref{fig:EBSlices}, we provide an example of snapshots of the fractional electric and magnetic field densities at three different times. The initial resonance instability leads to a growth of large length-scale modes with a somewhat larger fraction in electric fields. The third panel reveals a more scrambled configuration at late times (after backreaction). While we do not do so here, plotting the vector field configurations (rather than scalar energy densities), or pseudoscalar quantities such as $({\bf E}\cdot {\bf B})$ also provides useful insight into the complex underlying dynamics. 
\\ \\
\noindent{\it Energy and Gauss constraint preservation}: To keep track of the violation of the energy conservation in our simulations, we consider the quantity
\Beq
\label{eq:CurlyE}
\mathcal{E}\equiv \frac{|\mathcal{C}_{\mathcal{E}}|}{a^2\rho}=\left|1-\frac{3\mpl^2\mathcal{H}^2}{a^2\rho}\right|\,.
\Eeq
where $\mathcal{C}_{\mathcal{E}}$ was defined in section Eq.~\eqref{eq:FRWRych}. For the simulation whose results we have been discussing so far, the evolution of $\mathcal{E}$ is shown in the left panel in Fig. \ref{fig:NoVevConserv}. Note that it is easy to achieve a very small degree of energy violation, $<10^{-5}$, with a fairly large time step, due to the high order of the symplectic time integrator. Furthermore, the energy violation is quite stable and grows very slowly, due to the time-reversability of symplectic integrators.

We characterize the violation of the Gauss constraint with the quantity (see Eq. \eqref{eq:CGauss})
\Beq
\label{eq:CurlyG}
\mathcal{G}(\boldsymbol{\rm x})=\frac{e^{-2}|\mathcal{C}_{\rm G}(\tau,\textbf{x})|}{\sqrt{(\boldsymbol{\nabla}\cdot\boldsymbol{\pi}_{\!A}(\tau,\textbf{x}))^2+\left[\pi_1(\tau,\textbf{x})\varphi_2(\tau,\textbf{x})-\pi_2(\tau,\textbf{x})\varphi_1(\tau,\textbf{x})\right]^2}}\,.
\Eeq
We show its evolution at an arbitrary lattice point, $\boldsymbol{\rm x}$, in the right panel in Fig. \ref{fig:NoVevConserv}. The algorithm performance is excellent, with the violation never exceeding $10^{-6}$ and for most of the time remaining close to machine precision. The brief increase in the violation is observed only during the oscillatory phase, while the $\vp$ field is still homogeneous.  During this period, the calculation of spatial derivatives and/or differences of products of fields might lead to numerical errors due to attempting to compute small differences between large numbers, a phenomenon known as differencing noise. The observed growth in the violation during the oscillatory stage could be explained by such numerical errors.

\subsection{Nonlinear dynamics in the $v\ne 0$ case}
\begin{figure}[t]
\centering
  \includegraphics[width=5.8in]{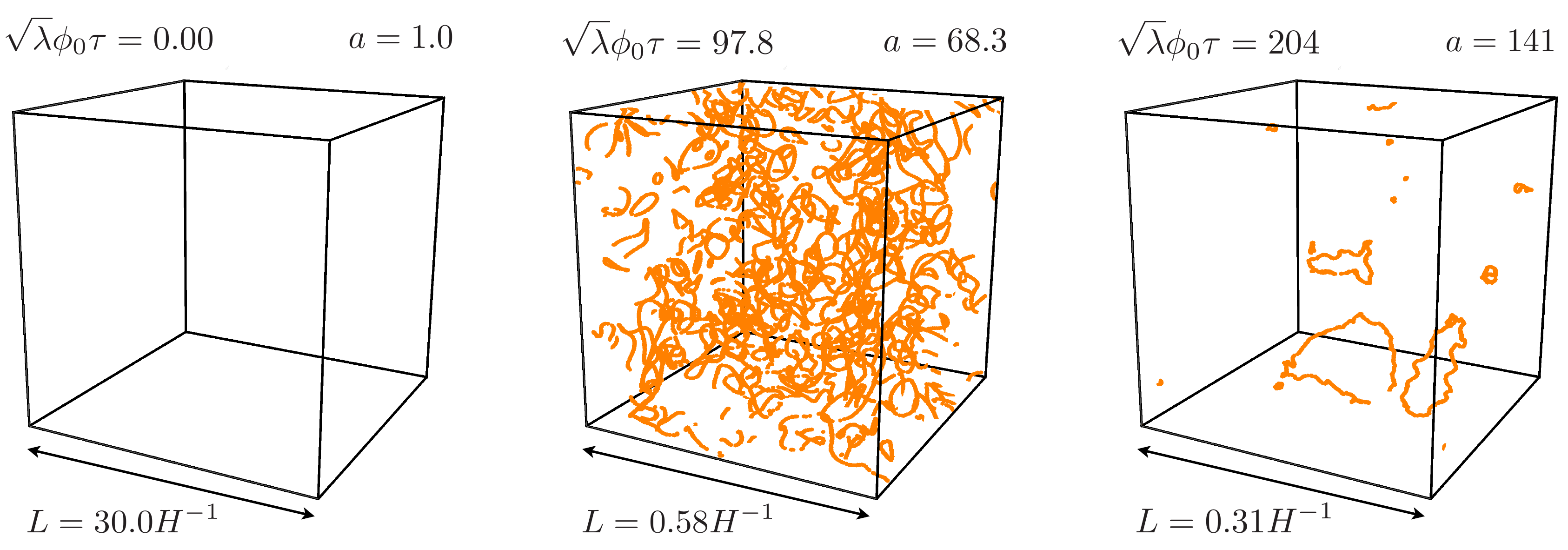} 
  \caption{Snapshots of the simulation box at four different times for the case where $V(\vp)$ has a Sombrero-hat shape. The orange points have non-zero winding number, $n$, see Eq. \eqref{eq:windingDefn}. The physical size of the simulation box, $L$, is given in units of the Hubble radius, $H^{-1}$. There is a copious production of subhorizon Nielsen-Olesen string loops around the time of backreaction. The loops eventually start to evaporate away. In the last panel the string core is resolved by $\mathcal{O}[10]$ points per linear dimension.}
  \label{fig:StringBox}
\end{figure}

We move on to the $v\neq0$ case, for which the profile of the Higgs  potential, $V(\vp)$, resembles a sombrero hat, see right panel in Fig. \ref{fig:PotBowl}. We set
\Beq
\label{eq:vvalue}
v=1.32\times 10^{-2} \mpl\,,
\Eeq
which is in agreement with Eq. \eqref{eq:Phi0v}. For the coupling constant $e$, we again use the value from Eq. \eqref{eq:elambda}.
\\ \\
{\it Field Power Spectra}: The initial evolution of the system proceeds in the same manner as in the previous case from Section \ref{sec:NoVevPreheating}. Since the $v$ from Eq. \eqref{eq:vvalue} is much less than the typical amplitude of $\bar{\varphi}_1$ oscillations, see Eq. \eqref{eq:Phi0v}, the initial parametric resonance phase is unaffected by $v$. We still have significant $\delta\boldsymbol{\rm A}$ resonant particle production. Again parametric resonance does not develop in the Higgs due to our choice of $e$, as explained in Section \ref{sec:NoVevPreheating}. Only once $\delta\boldsymbol{\rm A}$ begins to backreact, there is significant amplification of a broad range of comoving Higgs modes. After backreaction, the power spectra of the Higgs and the gauge fields again settle into stable broad single-peaked configurations. Since the power spectra plot are qualitatively similar to the $v=0$ case, we have relegated them to an appendix. 
\\ \\
\noindent{\it Cosmic strings}: Plotting the evolution of the fields in real space, reveals a phenomenon that cannot be picked out from the evolution of the power spectra. Note that the $v\neq0$ Higgs potential (right panel in Fig. \ref{fig:PotBowl}), can support the non-trivial field configurations known as topological strings \cite{1973NuPhB..61...45N}. They can be generated during thermal phase transitions via the Kibble mechanism in the form of cosmic string networks (for reviews see, e.g., \cite{Vilenkin:2000jqa,Hindmarsh:1994re,Copeland:2009ga,Drew:2019mzc}). Strings can be also produced after backreaction due to parametric resonance \cite{Felder:2001kt,Tkachev:1998dc,Kasuya:1997ha,Dufaux:2010cf}, just like in our case. Since strings are characterized by a non-zero integer topological number, known as the winding number, $n$,
\Beq
\label{eq:windingDefn}
n\equiv\frac{1}{2\pi}\oint d{\boldsymbol{l}}\cdot{\boldsymbol{\nabla}}\arg(\varphi)\,,
\Eeq
we plot the lattice points with $n\neq0$ at four different times in Fig. \ref{fig:StringBox}. 

The first panel in Fig. \ref{fig:StringBox} is at the start of the simulation. All lattice points have $n=0$, consistent with the inflationary initial conditions, see Eqs. \eqref{eq:InitBckgrnd} and \eqref{eq:InitFldsLat}. Towards the end of the resonant particle production and the onset of backreaction we observe copious formation of strings and string loops with a sub-Hubble correlation length, as shown in the second panel in Fig. \ref{fig:StringBox}. The strings then interact,\footnote{The 2-dimensional counterparts to our strings are known as vortices. The long-range interaction force between like-charged vortices is repulsive for $e^2<2\lambda$ \cite{Weinberg:2012pjx}, and hence for our parameter choice, Eq. \eqref{eq:elambda}. } reconnect into loops and gradually evaporate via classical radiation. We see features developing on loops, which split from the larger loop to form smaller loops, which then decay away. The last large loop in our simulation is seen (third panel in Fig. \ref{fig:StringBox}) around the time the fields enter the long-lived steady-state non-linear stage. At this stage, the string core is resolved by roughly 10 points per linear dimension. A more dedicated recent study of decay of single loops can be found in \cite{Matsunami:2019fss}. At late times, there are no strings in our box, which is consistent with the conservation of the net winding number and our initial conditions having zero $n$.\footnote{Another late-time configuration which is consistent with our initial conditions and the conservation of the topological charge is a pair of parallel strings with winding numbers of unlike signs (and therefore zero net winding number). We observed such final field configurations with both strings stretched across the same pair of opposite faces of the cubic box. The two strings were stationary with respect to the comoving lattice and not reconnecting into loops and evaporating for the duration of the simulations. The probability for such scenarios was quite low, $<10\%$, for an ensemble of initial field realizations, see Eq. \eqref{eq:stochasticInitNumbers}, and non-vanishing only for very small box sizes, much smaller than the Hubble radius around the time of backreaction and string formation.} We note that the string configurations never dominate the energy density in our box.
\begin{figure}[t]
\centering
\includegraphics[width=2.8in]{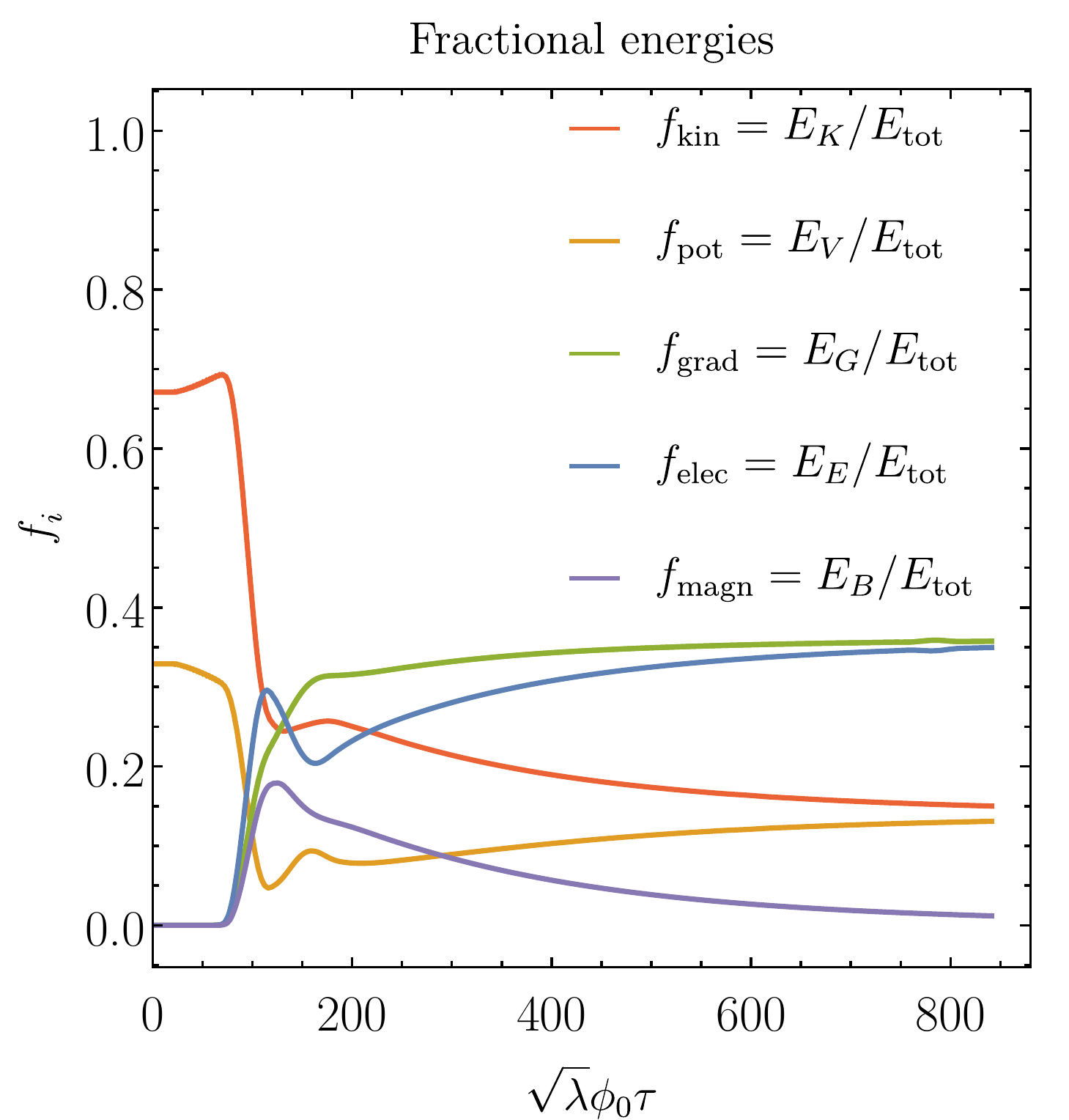}
\includegraphics[width=2.8in]{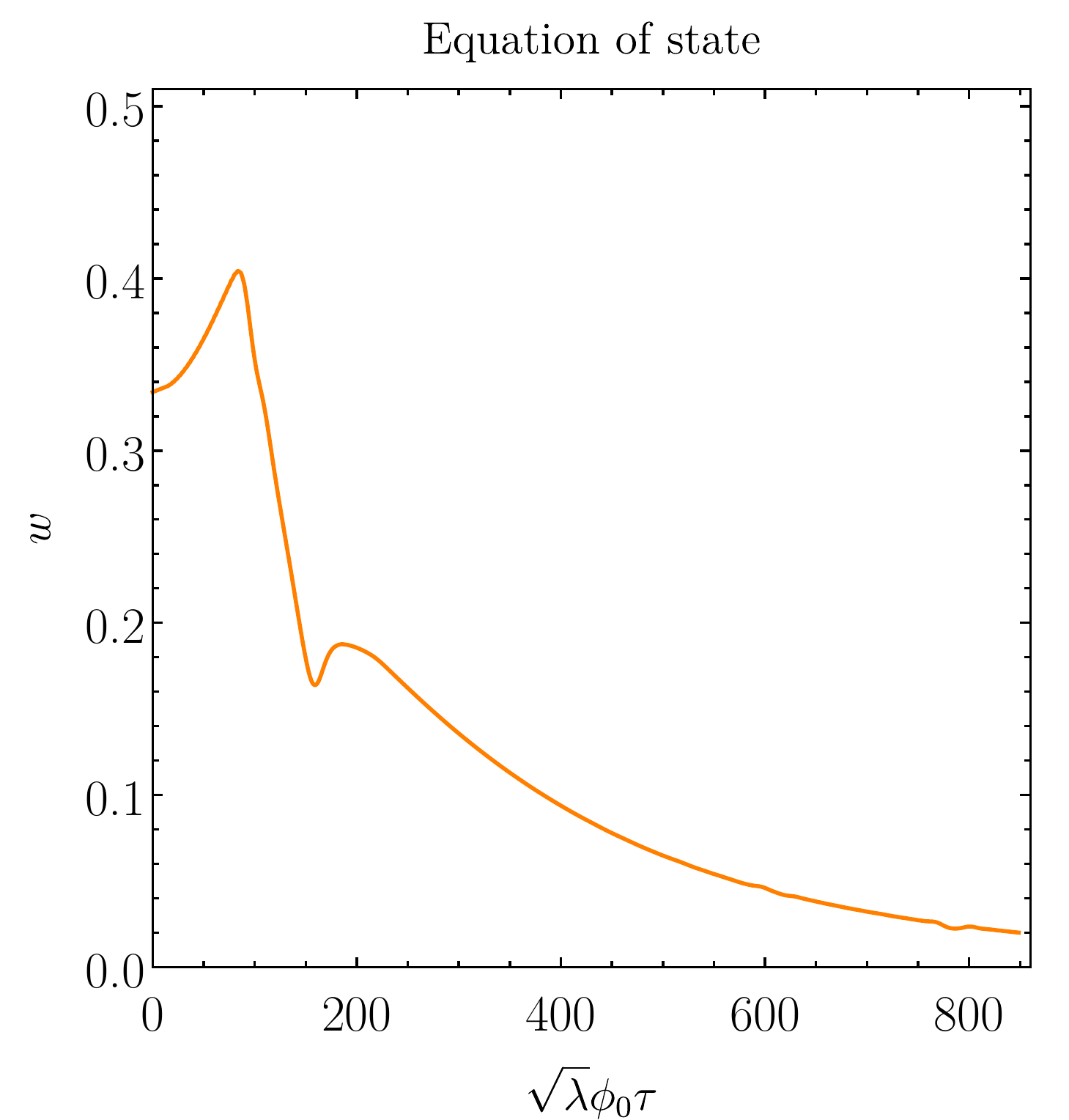}
\caption{Same as Fig. \ref{fig:NoVevEnergyFraction}, but for $v\neq0$. The initial oscillatory phase of homogeneous $\bar{\varphi}_1$ is still clearly visible in both panels. However, after backreaction, the complex Higgs falls into the potential valley at $|\varphi|=v/\sqrt{2}$, see Fig.~\ref{fig:PotBowl}. All dynamical components are now massive. The radial component of the Higgs (stuck near the bottom of potential valley) behaves as a massive scalar, as reflected by the red and orange curves in the left panel. The azimuthal component of the Higgs gets `eaten up' by the gauge field, rendering the latter massive, as depicted by the blue and green curves in the left panel. Note that $f_{\rm grad}$ includes the gauge field mass term from the Higgs mechanism. The magnetic energy gets diluted away, since it redshifts faster than the kinetic and potential energies of the massive Higgs and gauge fields, as shown by the purple curve in the left panel. The evolution of the equation of state (right panel) captures this transition from a radiation-like to a matter-like state of expansion.}
\label{fig:StringsEnergyFraction}
\end{figure}
\\ \\
\noindent{\it Energy fraction and equation of state evolution}: The evolution of the mean fractional energies, $f_i$, is shown in the left panel in Fig. \ref{fig:StringsEnergyFraction}. Just like in the case with zero $v$ from Section \ref{sec:NoVevPreheating}, we again have two distinct regimes with a brief transitionary period inbetween. During the initial oscillatory phase, most of the energy is stored in the oscillating $\bar{\varphi}_1$, again with $f_{\rm kin}\approx 2f_{\rm pot}$. This approximate equality (which is exact for a quartic Higgs potential) becomes less accurate with time, since the amplitude of $\bar{\varphi}_1\propto a^{-1}$ and the $v^2$ term in $V$ becomes increasingly important. Backreaction occurs around $\tau\sim 80(\sqrt{\lambda}\phi_0)^{-1}$ and is followed by a swift redistribution of energy. By $\tau\sim 200 (\sqrt{\lambda}\phi_0)^{-1}$, the last string loops start evaporating and the fields settle on a long and steady approach to equipartition. 

After this point, at almost every location on the lattice, the $\vp$ field has settled into the valley $V(|\vp|=v/\sqrt{2}$). The radial component of the $\vp$ field oscillates above the vev with amplitude $\ll v$ in a spatially inhomogeneous manner. Note that the radial Higgs component behaves as a massive scalar, since the bottom of the valley at $|\vp|=v/\sqrt{2}$ is quadratic. The azimuthal Higgs component (the massless Nambu-Goldstone boson) is `eaten up' by the gauge field like in the conventional Higgs mechanism, rendering the gauge field a massive vector field. Hence, both the radial Higgs degree of freedom and the gauge field behave as massive fields. They have massive non-relativistic modes, whose energy density redshifts as $\propto a^{-3}$, slower than the energy density of the relativistic modes (which scales as $\propto a^{-4}$). This implies that after a few {\it e}-folds of expansion, both the radial Higgs component and the gauge field should behave as pressureless matter, with potential and kinetic energies equal to each other and much greater than the gradient and curl energies. In our notation, this means that for the radial component of the Higgs, $f_{\rm kin}\rightarrow f_{\rm pot}$, whereas for the gauge field $f_{\rm elec}\rightarrow f_{\rm grad}$ (recall that the `potential' mass term for the gauge field is contained in $f_{\rm grad}$, see Eq. \eqref{eq:fisDefn}) and $f_{\rm magn}\rightarrow0$. Indeed, this is the late-time behaviour shown in the left panel in Fig. \ref{fig:StringsEnergyFraction}.
\begin{figure}[t]
 \centering
 \includegraphics[width=2.8in]{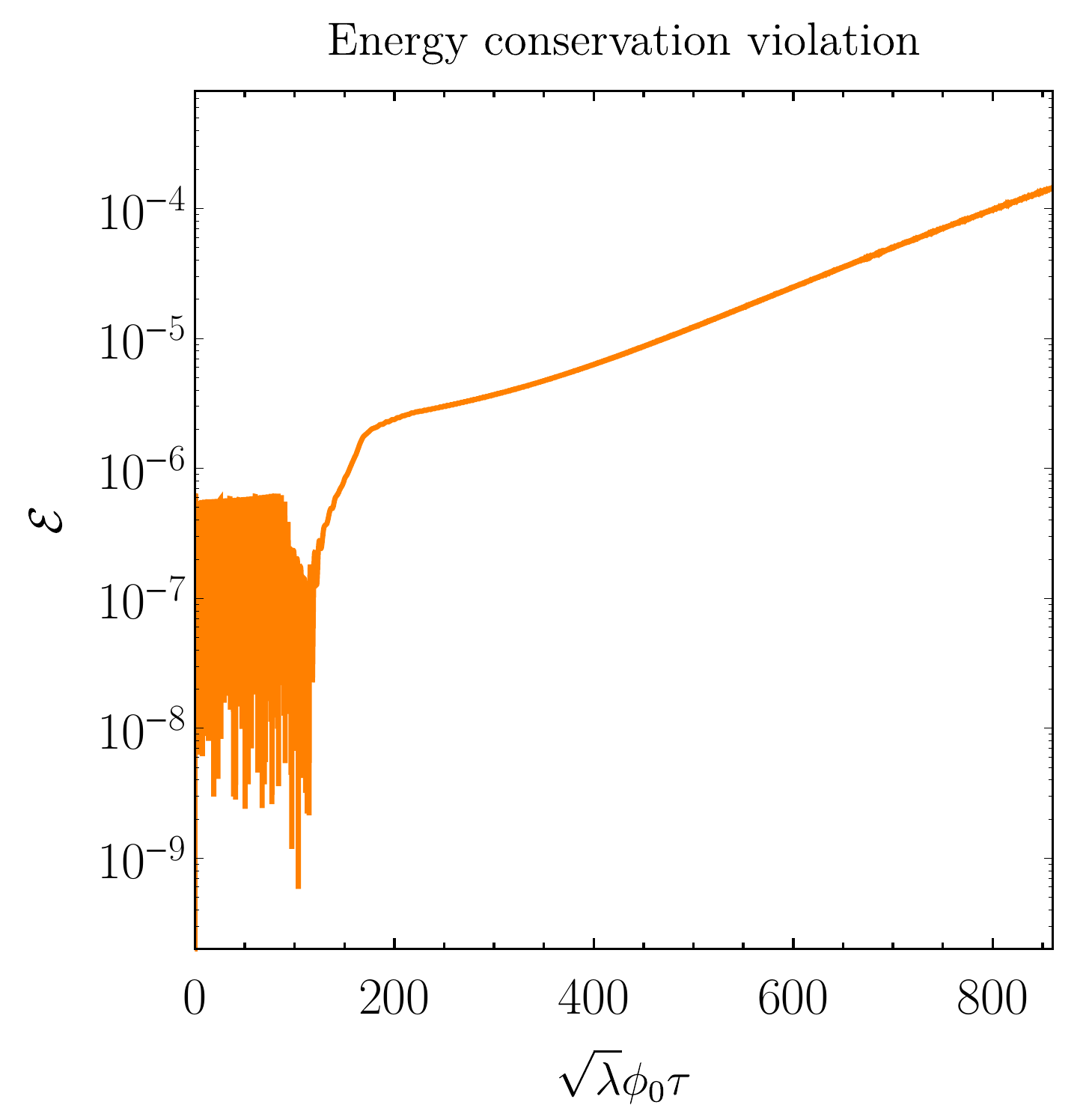}
 \includegraphics[width=2.8in]{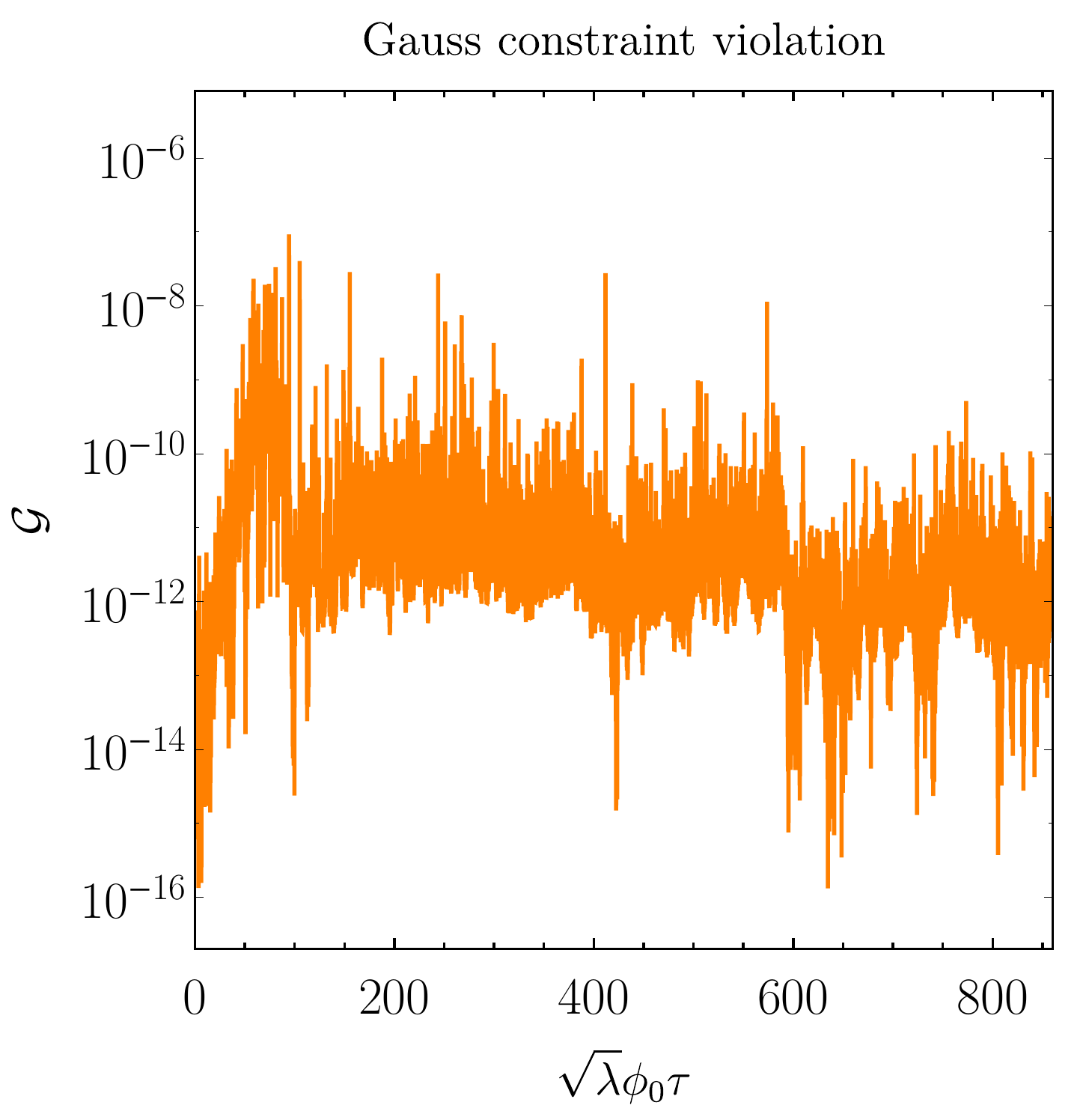}
 \caption{Same as Fig. \ref{fig:NoVevConserv}, but for $v\neq0$. The evolution of the violations of the energy conservation and Gauss constraint are qualitatively similar to the $v=0$ case. The slight difference is a small increase at late times of the energy violation only. It is due to the fact that we integrate in conformal time, whereas at late times the system is dominated by heavy, non-relativistic modes, see Fig. \ref{fig:StringsEnergyFraction}. For a possible improvement, see Fig. \ref{fig:StringsEnerConserv}.}
 \label{fig:StringsConserv}
\end{figure}

The self-consistent evolution of the mean equation of state, $w$ (see Eq. \eqref{eq:EoSDefn} for its definition) is given for the first time in the right panel in Fig. \ref{fig:StringsEnergyFraction}. Initially, it has a radiation-like value, due to the oscillations of the background $\bar{\varphi}_1$. The growing deviation from $1/3$ is due to the $v^2$ in $V$ becoming increasingly important as the amplitude of the oscillating $\bar{\varphi}_1$ is redshifted. After backreaction, $w$ steadily approaches $0$, since both the gauge field and the radial component of the complex Higgs field are massive. Their relativistic pressures are redshifted away, leaving behind pressureless scalar and vector dust with $w\rightarrow0$.
\\ \\
\noindent{\it Energy and Gauss Constraint}: The energy conservation is shown in the left panel in Fig. \ref{fig:StringsConserv}. We have used the same lattice parameters as for the simulation from Section \ref{sec:NoVevPreheating}. The energy conservation is still excellent, $\leq 10^{-4}$. It is almost identical to the one for the $v=0$ case given in the left panel in Fig. \ref{fig:NoVevConserv}, worsening only slightly at late times. The reason for this slightly worse performance for $v\neq0$ can be traced back to the fact that we work with a fixed conformal-time step, $\Delta\tau$. For $v=0$, i.e., a quartic Higgs potential, the typical frequency scales always decrease with time as $\propto a^{-1}$, implying that their product with the cosmic-time step, $a(\tau)\Delta\tau$, is constant. 

\begin{figure}[t]
\centering
\includegraphics[width=2.8in]{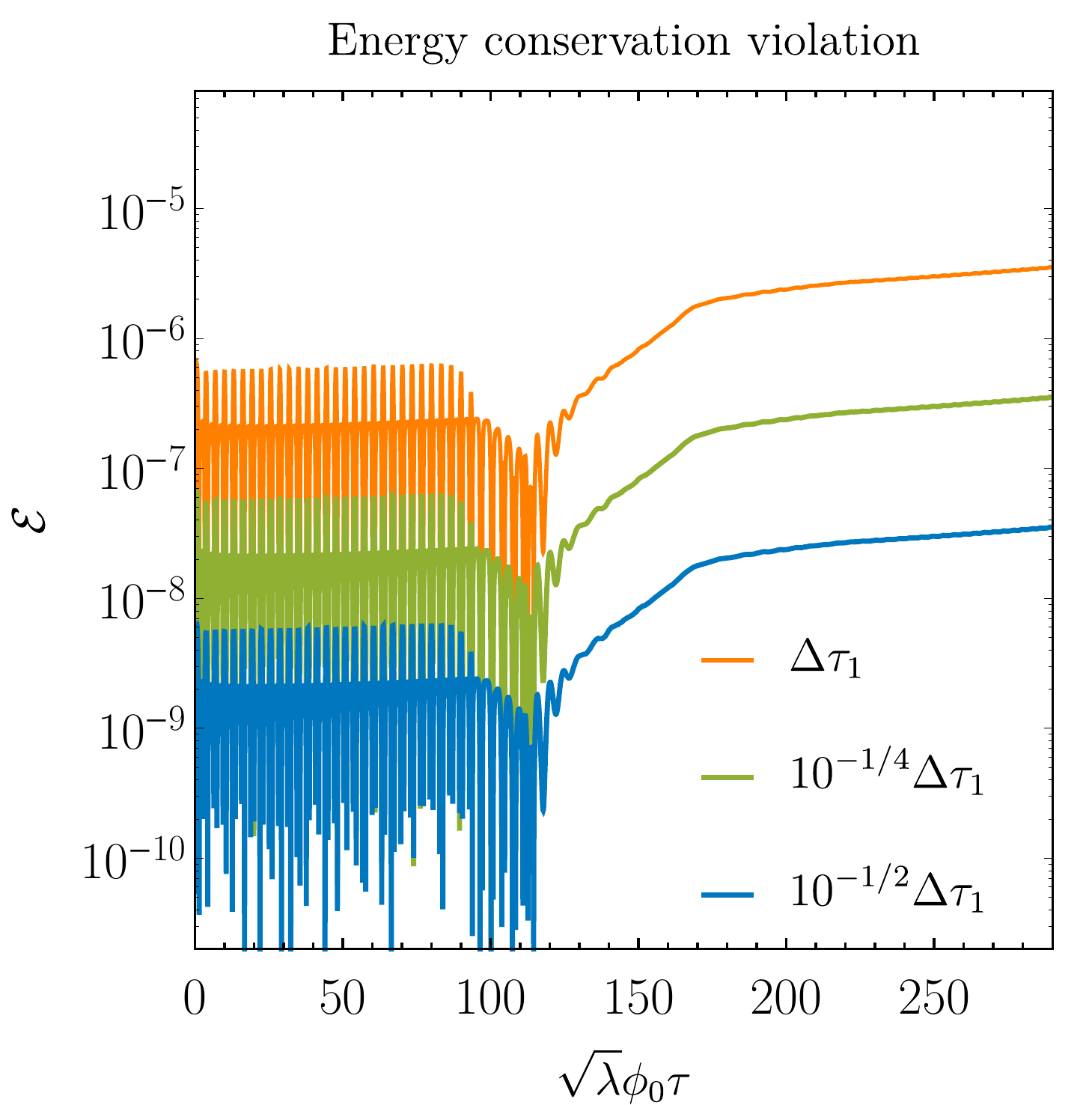}
\includegraphics[width=2.8in]{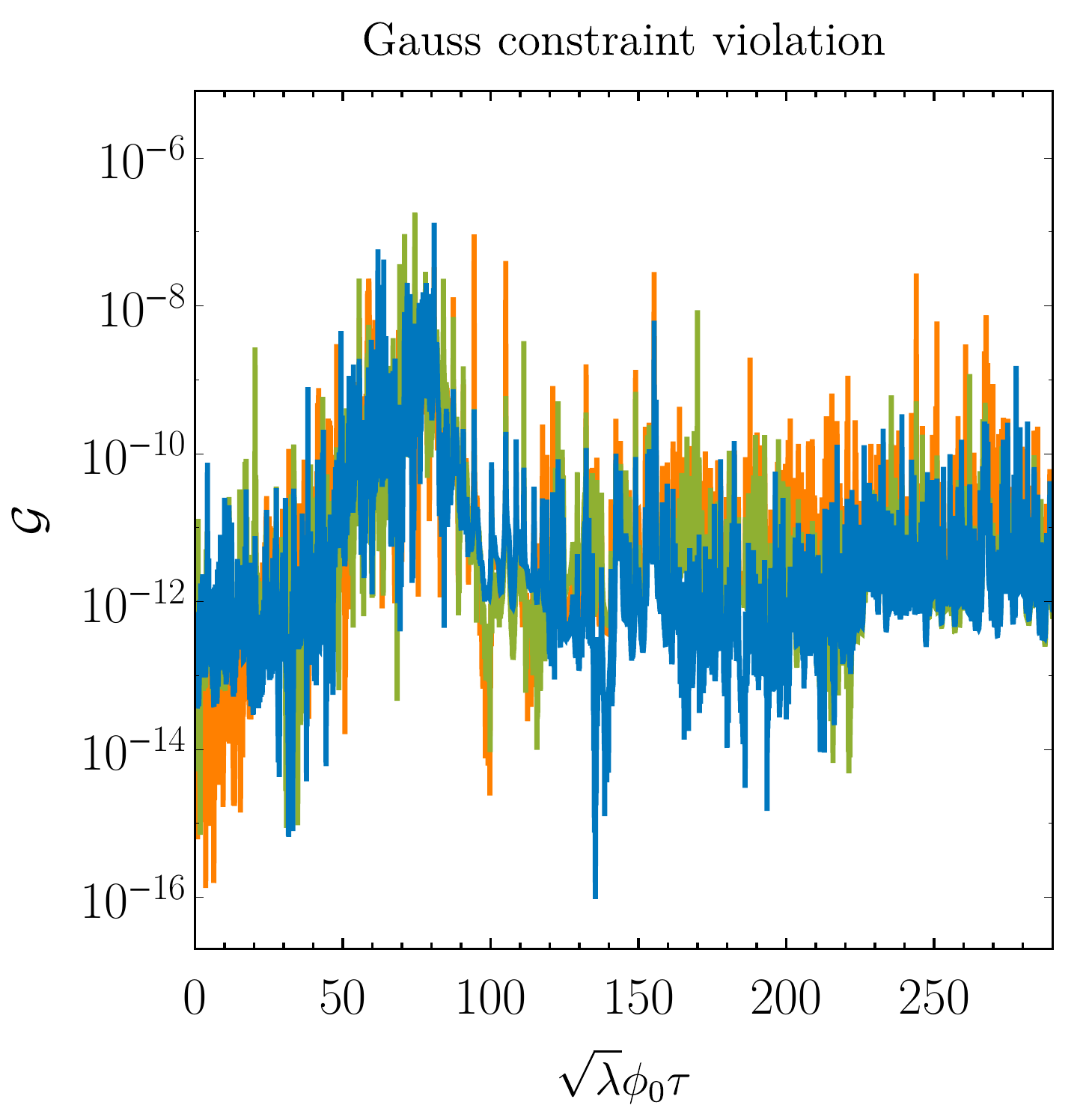}
\caption{{\it Left panel}:The energy conservation for three different conformal time steps. The red curve is for the case from the left panel in Fig. \ref{fig:StringsConserv}. The green and blue curves are for the same simulation and model parameters, but for time steps $10^{1/4}$ and $10^{1/2}$ times smaller. The used symplectic integrator was of fourth order, $k=4$, and the energy conservation scales appropriately with the time step, $\propto\mathcal{O}(\Delta\tau^k)$. {\it Right panel}: As expected from our algorithm, the violation of the Gauss constraint does not depend on the size of the time step.}
\label{fig:StringsEnerConserv}
\end{figure}

On the other hand, for the massive case, $v\neq0$, the typical frequency scales are constant, implying that their product with the cosmic-time step grows like $\propto a(\tau)$, thereby increasing the time-integration error. Even though it was not necessary for this study, this small degradation in energy conservation can be easily alleviated by decreasing the conformal-time step only slightly. This takes advantage of the fact that the order of the time integrator, $k$, is high and the energy conservation is quite sensitive to the time step. The local truncation error in the time integration is $\mathcal{O}(\Delta\tau^{k+1})$, see Eq. \eqref{eq:Kk}, and the total accumulated error is $\mathcal{O}(\Delta\tau^{k})$. For $k=4$, the energy conservation can be improved by one or two orders of magnitude, when we decrease the conformal-time step only by a factor of $10^{1/4}$ or $10^{1/2}$, respectively, as shown in Fig. \ref{fig:StringsEnerConserv}. 
\\ \\
The Gauss constraint violation is shown in the right panel in Fig. \ref{fig:StringsConserv} for a random lattice point. We find that it is qualitatively identical to the one for the $v=0$ case given in Fig. \ref{fig:NoVevConserv}. It is also insensitive to the conformal-time step, which is expected for a quantity dominated by differencing noise.

\section{Discussion}
\label{sec:Disc}

We have presented a novel prescription for  numerically evolving Scalar Electrodynamics in FRW spacetime (sometimes also referred to Abelian-Higgs system, or explicitly, a charged scalar minimally coupled to Abelian gauge fields) . Our prescription combines two different well-known techniques in numerical simulations of related systems,  one for handling the spatial derivatives, and another for temporal derivatives. The spatial discretization is carried out according to the Lattice Gauge Field theory prescription (using the lattice links formalism). The time evolution is performed with symplectic integrators. The algorithm allows for the self-consistent evolution of the FRW scale factor and fields, while respecting the Gauss constraint exactly on the discretized lattice. 
To the best of our knowledge, this is the first explicit-in-time algorithm which guarantees the preservation of of the Gauss constraint, while solving for the expansion of the universe self-consistently. In addition, the time integrator can be made of arbitrary high order, without violating any of the algorithm's properties.\footnote{The exact preservation at the level of the algorithm is violated in an actual numerical calculation due to, for example, differencing noise.} 

The use of symplectic integrators was inspired by already existing scalar field lattice codes such as {\sc HLattice} \cite{Huang:2011gf}, {\sc Defrost} \cite{Frolov:2008hy} and {\sc PyCool} \cite{Sainio:2012mw}. They are known for their excellent energy conservation and stability, when solving for the expansion of the universe self-consistently. Perhaps, one of the reasons why such integrators have not been employed in gauge field studies is the uncertainty of whether they will respect the Gauss constraint equation -- a non-dynamical equation which always appears in gauge field theories, but never in pure-scalar models. 

The issue with the Gauss constraint equation has long been solved in flat spacetime, by discretizing the Higgs field on a 4d rigid lattice, and defining the gauge fields on the lattice links. This method has been successfully extended to FRW spacetimes, with the scale factor evolving according to some fixed power-law, e.g., $a(\tau)\propto \tau$, not determined by the evolution of the Higgs and gauge fields \cite{Dufaux:2010cf,Figueroa:2015rqa}.\footnote{A more detailed discussion of how our work fits in the context of earlier literature was provided in the introduction.}

Given these successes, we employed a combination of the two approaches to achieve both good self-consistent expansion and preservation of the Gauss constraint. We showed how to combine them in a very straightforward way. To use symplectic integrators, we needed a simple Hamiltonian. The Hamiltonian for Scalar Electrodynamics was simplified substantially by working in the $A_0=0$ gauge. This gauge choice made all kinetic terms canonical, which in turn gave use a symplectic integrator for the time evolution. The remaining spatial derivatives were discretized by defining the complex scalar and the 3-components of the gauge field on a 3d spatial lattice, similarly to Lattice Gauge Field theory (with the use of Link variables). This automatically allowed the system to respect the residual gauge freedom in $A_0=0$ gauge. More importantly, this spatial discretization yielded a well-defined expression for the lattice version of the Gauss constraint, consistent with the residual symmetries. This version of the Gauss constraint turned out to be respected exactly by the symplectic integrators.
\\ \\
\noindent {\it Numerical Studies}: To test our algorithm and code, we investigated two different reheating scenarios. In both, the role of the oscillating inflaton was played by the real component of the complex scalar field. In the first scenario, the scalar potential was a simple quartic minimum (i.e., `unbroken'), whereas in the second one, the scalar field potential allowed for spontaneous symmetry breaking (a Higgs-like potential). The lattice simulations captured the initial preheating phase in which the gauge field was excited non-perturbatively due to parametric resonance in the oscillating homogeneous scalar field background. They also revealed the  subsequent non-linear stage after backreaction and fragmentation of the condensate.  All qualitative predictions of linear analyses for the resonant particle production were reproduced. The subsequent non-linear stage included interesting non-linear phenomena such as the formation and disappearance of  Nielsen-Olesen strings (in the case with the symmetry breaking potential). 

The self-consistent FRW expansion was computed throughout -- the energy conservation violation was stable, being always $\leq 10^{-4}$ and scaling appropriately with the size of the time step, $\propto \Delta\tau^k$ for a $k$th-order integrator. As expected, the case with unbroken scalar field potential lead to a radiation-like equation of state during the oscillatory as well as the non-linear stages. Similarly, a late-time matter-like state of expansion  predicted for the spontaneously broken case (since along with the radial part of the scalar, the gauge fields are now massive), was reproduced during the non-linear stage. In both cases the Gauss constraint was preserved, with violations $<10^{-6}$; these violations were dominated by numerical errors due to subtraction of large quantities with very small differences (differencing noise).
\\ \\
{\noindent \it Limitations}:
We note that the combination of a symplectic time integrator and a Lattice Gauge Field type of discretization does not seem to work for non-Abelian gauge theories, e.g., an $SU(2)$ Yang-Mills theory with a Higgs doublet. In this case, the Gauss constraint is not respected by the symplectic integrator. A reason for that could be the non-linear nature of the Gauss constraint in non-Abelian theories. Another class of gauge field models which is not well-suited for our prescription is the one featuring an axion, $\chi$. Both Abelian and non-Abelian theories with a Chern-Simons type of interaction, $\propto\chi F\tilde{F}$, cannot be integrated symplectically due to the non-canonical structure of the gauge field kinetic term. However, an interaction of the form $\propto \chi F^2$ could work at least in Abelian theories, since the kinetic term of the gauge field is canonical up to an axion-dependent rescaling, meeting the conditions for usage of symplectic integrators. 
\\ \\ 
\noindent We hope that ${\sf GFiRe}$ will be used for many more cosmological studies of non-linear gauge field dynamics. We plan to make ${\sf GFiRe}$ public in the near future.

\acknowledgments

We thank Andrei Frolov, Andrew Long, Ed Copeland, Eiichiro Komatsu, Jonathan Braden, Matthew Reece, Paul Saffin and Peter Adshead  for useful comments and discussions. MA is supported by a DOE grant DE-SC0018216. Part of this work was carried out at the Aspen Center for Physics, which is supported by National Science Foundation grant PHY-1607611.

\bibliography{GFiRe}
\bibliographystyle{JHEP}
\newpage
\appendix
\section{Power Spectra for $v\ne 0$ case}
We provide the time evolution of the power spectra for the gauge fields as well as the real and imaginary components of the $\vp$ field below in the $v\ne0$ case. See corresponding discussion in the main text. 
\begin{figure}[!h]
\centering
 \includegraphics[width=3.0in]{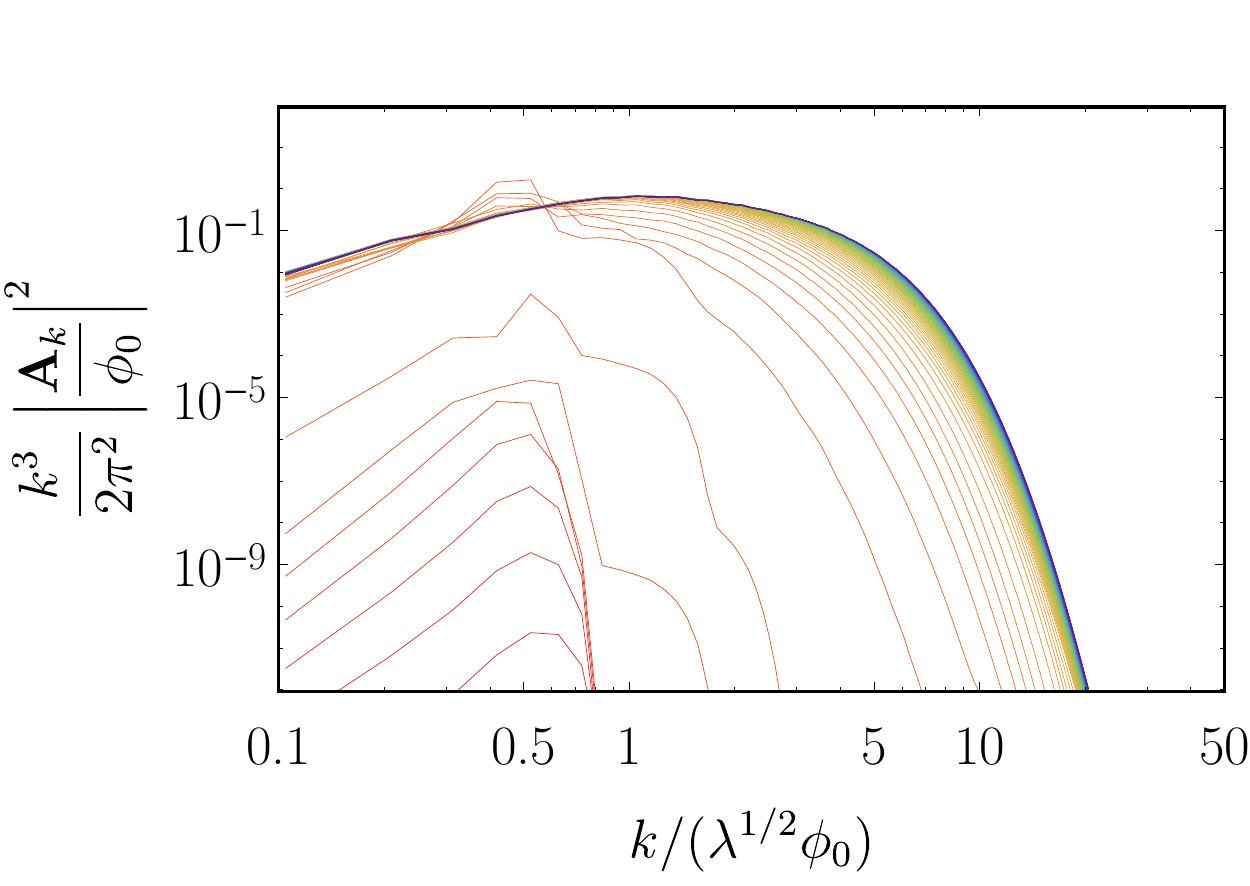}\\
   \includegraphics[width=3.0in]{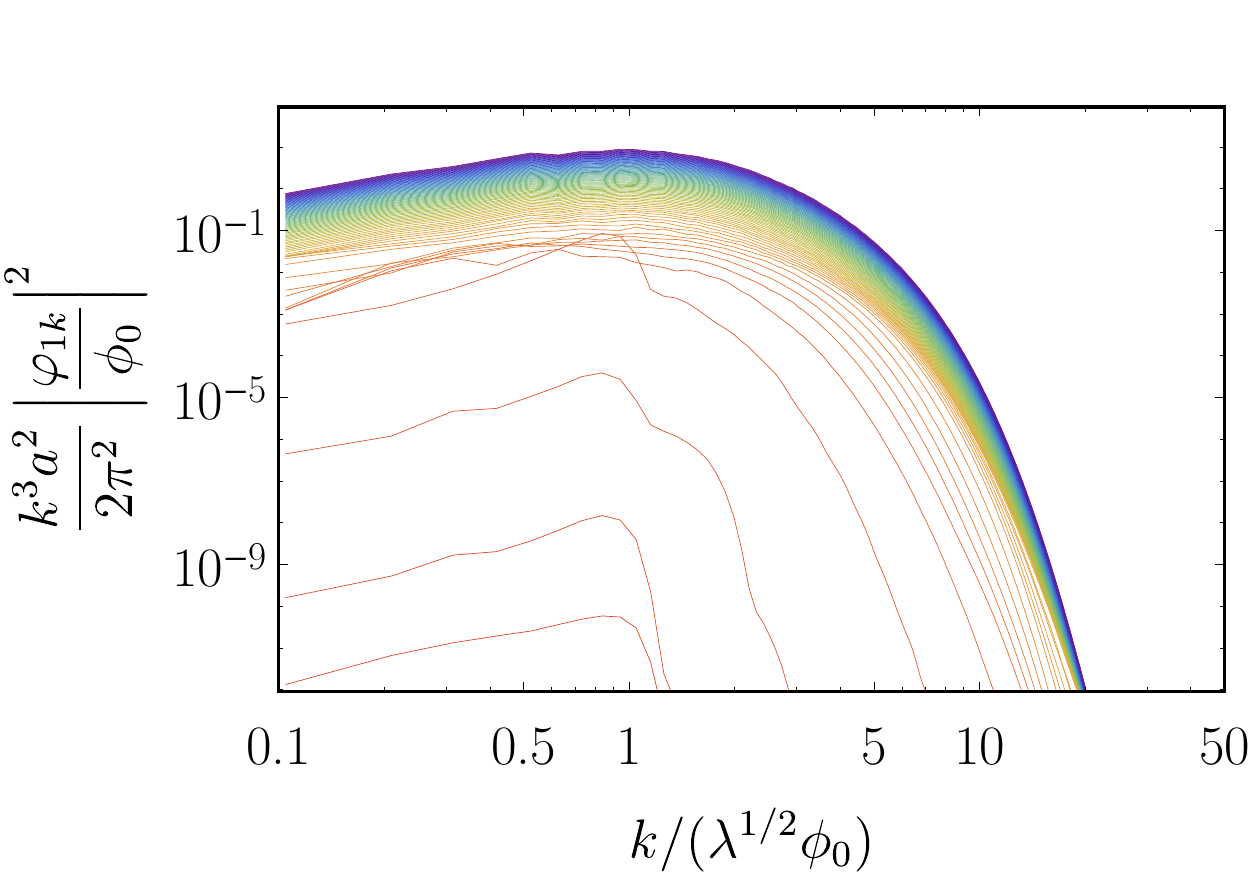}\\
   \includegraphics[width=3.0in]{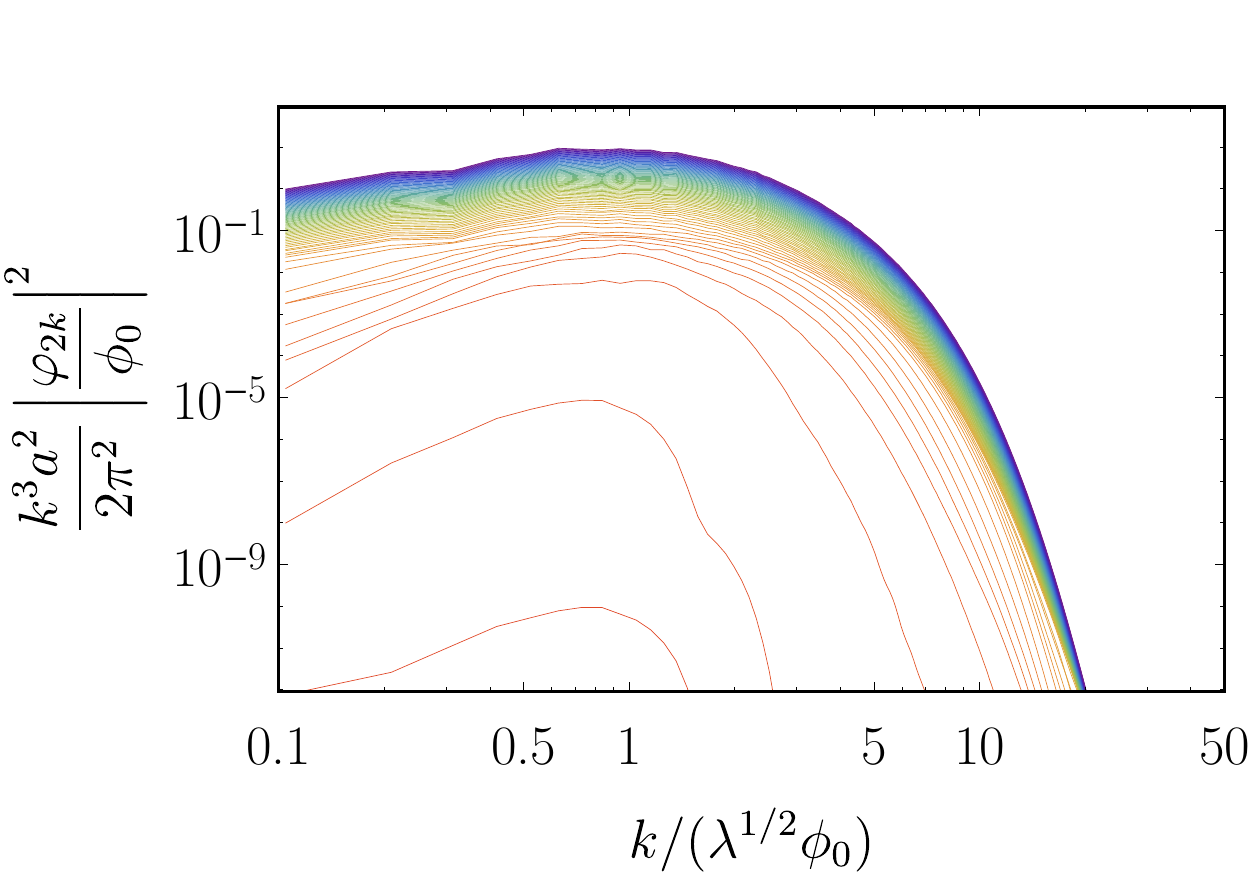}
   \caption{The evolution of the power spectra is qualitatively the same as in the $v=0$ case.}
   \label{fig:pspStringsPhi}
\end{figure}

\end{document}